\documentclass[final, 11pt, 3p]{elsarticle}

\usepackage{amssymb}
\usepackage{amsmath}
\usepackage{booktabs}
\usepackage{multirow}
\usepackage{url}
\usepackage{makecell}
\usepackage{float}
\usepackage{rotating}
\usepackage{setspace}
\usepackage[markup=underlined]{changes}

\journal{Transportation Research Part D: Transport and Environment}

\begin{document}
\doublespacing

\begin{frontmatter}



\title{Understanding Demand for Shared Autonomous Micro-Mobility} 


\author[inst1]{Naroa Coretti Sanchez}
\author[inst1]{Kent Larson}

\affiliation[inst1]{organization={MIT Media Lab},
            addressline={20 Ames Street}, 
            city={Cambridge},
            postcode={02142}, 
            state={Massachusetts},
            country={United States}}
\begin{abstract}
This study examines the behavioral and environmental implications of shared autonomous micro-mobility systems, focusing on autonomous bicycles and their integration with transit in the U.S. While prior research has addressed operational and lifecycle aspects, a critical gap remains in understanding which modes these services are likely to substitute, who is most inclined to adopt them, and how service attributes influence user decisions. We design a context-aware stated preference survey grounded in real-world trips and estimate discrete choice models, including a hybrid model incorporating latent attitudes. Findings indicate that adoption, mode shift, and environmental impacts are highly sensitive to service design. Scenarios with minimal wait and cost yield high adoption but increase emissions, while moderate waits are more likely to reduce impacts.  Adoption likelihood varies with demographic characteristics, and outcomes depend on city type, context, and infrastructure assumptions. These insights can inform the development of more sustainable and equitable mobility systems.

\end{abstract}



\end{frontmatter}

\section{Introduction}\label{sec:Introdution}


The concept of shared autonomous micro-mobility has recently emerged as an alternative that brings together autonomous driving technology with walkable city principles that aim to reduce car dependency \citep{sanchez2024shared}. While most AV development has centered on cars and taxis, shared autonomous micro-mobility introduces smaller vehicles based on the idea that the same qualities that make AVs appealing, such as convenience, efficiency, and flexibility, can also make sustainable mobility options more attractive.

Shared autonomous micro-mobility vehicles are ultra-lightweight autonomous vehicles such as autonomous bicycles, tricycles, and scooters, designed for single-person, low-speed operation. They operate as a shared on-demand service, where trips would occur as follows: When a user requests a ride, an autonomous vehicle would autonomously travel to the user's location. Then, the user can ride it just as a regular micro-mobility vehicle. Upon arrival at the destination, the vehicle would transition back to autonomous mode and proceeds to either find a charging station or to pick up the following user. \cite{sanchez2020autonomous}. In such a system, users would enjoy the convenience of not needing to search for available vehicles or return them to fixed docking stations, as the vehicles reposition themselves autonomously. This also eliminates the need for manual rebalancing operations, which are both costly and environmentally burdensome in current shared micromobility systems \cite{sanchez2024shared}.

Shared autonomous micro-mobility vehicles are actively being developed by both industry and academia \citep{sanchez2020autonomous,lin2021affordable, Tortoise, TechCrunch, manoeva2022investigation} and a growing body of literature has emerged examining the fleet-level dynamics of these systems, focusing on key aspects such as operational performance, service quality, and environmental sustainability \citep{sanchez2022performance, sanchez2022simulation, sanchez2022can, kondor2021estimating, genua2024shared}. Studies to date have identified several potential benefits of shared autonomous micro-mobility systems. These include increased vehicle utilization and operational efficiency through automation \citep{kondor2021estimating}, reduced fleet sizes compared to traditional bike-sharing systems \citep{sanchez2022performance}, and lower environmental impacts by eliminating the need for rebalancing with vans or trucks \citep{sanchez2022can}. Additional research has highlighted potential applications beyond passenger mobility—such as food delivery and multifunctional urban services—further improving resource efficiency \citep{genua2024shared, coretti2024multifunctional}. Decentralized fleet management strategies have also been proposed to support system scalability without relying on centralized infrastructure \citep{coretti2023urban, alfeo2019urban}. Collectively, these findings suggest that shared autonomous micro-mobility systems could serve as a flexible, lower-impact alternative to conventional urban mobility systems. However, the benefits are also shown to be highly dependent on scenario, system design, and local context, underscoring the need for further research.

In particular, while operational performance and environmental impacts of shared autonomous micro-mobility systems have been explored, limited attention has been given to understanding individual-level demand dynamics, including the mode shift they would generate and who the potential users would be. Understanding such patterns is critical for evaluating their broader environmental and societal implications. For instance, early studies on bike-sharing systems were conducted only after widespread deployment, revealing unexpected outcomes of often substituting a higher share of walking trips rather than car trips, which limits their sustainability potential \citep{teixeira2021empirical, de2021environmental}. This highlights the necessity of studying mode choice behavior before large-scale deployment of new mobility technologies.

While modeling demand and sustainability implications for emerging modes comes with inherent challenges due to the lack of historical usage data, the pre-deployment stage offers a unique window of opportunity for research and policy development \citep{bergerson2020life}. Early-stage insights can inform vehicle design, business models, and urban planning frameworks. In this context, proactive demand modeling can help align stakeholder goals and support evidence-based policy-making to ensure sustainable and equitable service deployment. 

With the goal of understanding demand dynamics of shared autonomous micro-mobility vehicles, this paper presents a choice modeling study based on a context-aware, pivot-style stated preference survey conducted in the U.S. Based on this data, we estimate a set of discrete choice models, including a hybrid choice model that incorporates latent attitudes.  This approach allows us to explore mode substitution patterns, potential adoption rates, and environmental implications of deploying shared autonomous micro-mobility systems in different urban environments and with different wait time and cost levels. 

The remainder of the paper is structured as follows. Section \ref{sec:Previouswork} reviews relevant literature on shared autonomous micro-mobility systems and demand modeling for emerging mobility modes. Section \ref{sec:surveyandData} outlines the survey design, data collection procedures, and sample bias correction methods. Section \ref{sec:modeling} details the model development, estimation, and selection processes. Section \ref{sec:Results} presents the main findings, including behavioral insights, sociodemographic patterns, predicted mode shares, mode shift dynamics, and environmental impacts. Section \ref{sec:discussion} discusses the main findings, while \ref{sec:limitations} lists some of the main limitations of this study. Lastly, Section \ref{sec:Conclusion} offers concluding remarks and discusses broader implications.

\section{Previous work}\label{sec:Previouswork}

\subsection{Modeling the Performance and Implications of Shared Autonomous Micro-Mobility}

Several studies have explored the fleet-level performance of shared autonomous micro-mobility systems. For instance, \citet{sanchez2022simulation} developed a simulation model for analyzing shared autonomous micro-mobility systems in urban settings. Building on this work, \citet{sanchez2022performance} applied the model to a case study in Boston, USA, finding significant fleet size reductions compared to station-based and dockless shared bicycle fleets. Similarly, \citet{kondor2021estimating} evaluated the operational efficiency of shared autonomous scooters, showing that their utilization rates could be remarkably higher than current shared bicycle systems, indicating substantial performance improvements through automation.

Beyond passenger mobility, shared autonomous micro-mobility systems have been proposed for other urban functions. \citet{genua2024shared} assessed the performance of lightweight autonomous vehicles for food deliveries, finding that they could reduce both fleet sizes and environmental impacts compared to car-based delivery systems. Expanding on this, \citet{coretti2024multifunctional} explored multi-functional vehicle fleets that would cater to both on-demand rides and food deliveries, demonstrating that such vehicles could reduce fleet size requirements compared to having single-use dedicated fleets.

In terms of economic feasibility, \citet{salah2024could} investigated the financial sustainability of autonomous shared bicycles, highlighting the importance of balancing system costs with user adoption rates, service pricing, and fleet maintenance efficiency.

System scalability and operational robustness have received attention through decentralized system designs. \citet{alfeo2019urban} proposed a bio-inspired algorithm for shared autonomous micro-mobility vehicles to collect urban waste in a self-organizing, decentralized way. Similarly, \citet{coretti2023urban} developed a decentralized vehicle rebalancing algorithm for shared autonomous micro-mobility systems, demonstrating its potential to optimize vehicle distribution dynamically without relying on centralized control systems.

From an environmental perspective, \citet{sanchez2022can} found that autonomous bicycles could reduce $CO_2,eq$ emissions per passenger kilometer compared to current station-based and dockless bike-sharing systems, thanks to the smaller fleet sizes required and the elimination of the need to rebalance the vehicles with vans or trucks.  However, their results also show that the overall environmental impact of shared autonomous micro-mobility systems ultimately depends on the mode shift they generate. Results calculated using mode-shift data from nineteen different bicycle-sharing systems showed that overall environmental impact changes could range from a -43.3\% to +137.9\%. These findings underscore the need to design shared autonomous micro-mobility systems that maximize mode shifts away from polluting modes (e.g., cars and taxis) while creating positive synergies with more sustainable modes such as walking, biking, and public transit.  It must be noted that the results reported by \citet{sanchez2022can} are based on existing bicycle-sharing systems. However,  current bike-sharing demand patterns may differ from shared autonomous micro-mobility  due to its mobility-on-demand nature. The unique characteristics of shared autonomous micro-mobility could result in distinct mode-shift patterns, which highlight the need for covering the current gap in shared autonomous micro-mobility demand modeling.

\subsection{Demand Modeling Methods for Emerging Mobility Modes} 

Estimating demand for emerging mobility modes presents a significant challenge due to the lack of real-world historical data, which limits the availability of revealed preference (RP) observations. In such contexts, researchers commonly rely on stated preference (SP) surveys, where respondents are presented with hypothetical travel scenarios involving new mobility services. This approach allows for exploring potential decision-making processes in the absence of actual system deployment.

SP surveys generate valuable behavioral data by simulating scenarios that reflect key service attributes, such as travel time, cost, and availability. They have been widely used to estimate demand for micro-mobility, on-demand services, and autonomous vehicles \citep{asgari2020incorporating,esztergar2022assessment, frei2017flexing,sweet2021user,jiang2019capturing}. However, SP surveys are susceptible to hypothetical bias, as individuals may respond differently in hypothetical scenarios than they would in real-world settings. 

To mitigate this issue, pivot-style and context-aware SP surveys are frequently employed, where hypothetical scenarios are generated by adjusting key attributes of a respondent's recent real-world trip \citep{train2008estimation, rose2009dual, danaf2019context}. This approach enhances behavioral realism by grounding hypothetical choices in familiar contexts. For instance, \citet{danaf2019context} estimated the demand for a new on-demand mobility service by building SP scenarios derived from RP data collected via smartphone tracking. 

In parallel, hybrid choice models (HCMs), also known as Integrated Choice and Latent Variable (ICLV) models, have gained prominence for incorporating latent psychological constructs—such as attitudes or perceptions—into discrete choice frameworks \citep{abou2024hybrid}. By explicitly modeling these unobserved factors, HCMs provide a richer behavioral representation, which has made them popular for cycling choice modeling \citep{scorrano2021active}.

As discussed in the previous section, understanding the demand for shared autonomous micro-mobility is essential for assessing its potential impacts on sustainability, travel behavior, and system viability. Yet, demand modeling in this domain remains at an early stage.  \cite{kania2023data} define aggregate demand scenarios for autonomous cargo bike fleets based on survey responses, which they aggregate to estimate overall usage potential. While this method provides a useful approximation of potential usage, it does not incorporate a discrete choice framework and therefore lacks a behavioral basis for modeling individual decision-making processes. Consequently, it cannot account for trade-offs between competing alternatives or simulate responses to changes in service attributes such as cost or wait time—factors that are essential for policy evaluation and system design.

\cite{zou2024bike} take a step further by estimating discrete choice models based on stated preference data for autonomous scooter adoption. Their results demonstrate that both the ability to summon a vehicle (autonomy) and the presence of bike lanes significantly increase the likelihood of adoption. While this work contributes important behavioral insights, it does not extend to estimating mode shares or modeling mode shifts from existing travel modes, nor does it assess the broader environmental implications of such shifts. 

The present study extends this body of work by developing a series of discrete choice models to analyze individual-level adoption of shared autonomous micro-mobility alternatives. It focuses on autonomous bicycles and their integration with public transit, and builds on the aforementioned methodological developments by employing a context-aware SP design and testing an HCM to investigate adoption patterns. Unlike prior studies, this work emphasizes the estimation of mode shares and mode shift dynamics, capturing how individuals make trade-offs across cost, wait time, and multimodal configurations. These demand estimates serve as the foundation for evaluating system-level outcomes, including potential environmental impacts and demographics, and thereby support more informed policy and planning decisions regarding the deployment of shared autonomous micro-mobility systems.





\section{Survey design and data collection}\label{sec:surveyandData}

This section describes the survey design and data collection processes, which form the basis for the estimation of a set of Discrete Choice Models (DCMs) detailed in Section \ref{sec:modeling}.

The survey consisted of three main components: (1) a revealed preference section where respondents recalled a recent trip using their usual commuting mode, (2) a stated preference section with six hypothetical mode choice tasks, and (3) a set of attitudinal questions.

In the stated preference tasks, respondents chose between three alternatives: (i) Autonomous Bicycles, (ii) Autonomous Bicycles combined with Public Transit, and (iii) their original mode. Each task varied the autonomous options across two attributes: cost and wait time. Each respondent answered 30 questions, including consent, eligibility filters, trip entry, questions on the availability of other modes, six stated preference tasks, two attention checks, 11 attitudinal items, and an optional feedback prompt. The survey was administered to 1,925 U.S.-based respondents, of whom 730 were retained as the final sample after applying pre-screening criteria and trip-based quality filters  (see Table \ref{tab:respondent_summary}).

The following subsections outline the structure of the survey, the data collection methodology, and the procedures used to correct for sample bias.

\subsection{Survey Design} \label{sec:survey-design}

\begin{figure}[!htb]
  \centering
  \includegraphics[width=0.7\columnwidth]{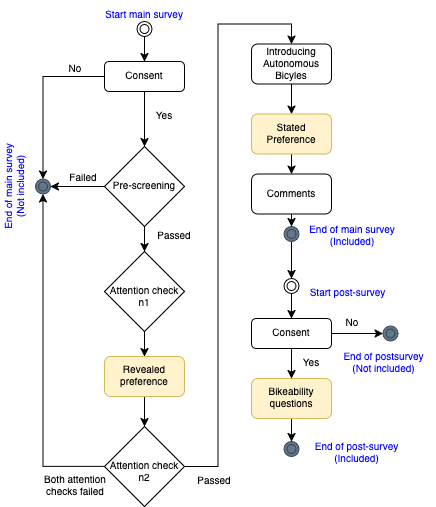}
    \caption{Survey flowchart detailing respondent progression through the main survey and the post-survey. The main survey included consent, pre-screening, two attention checks, a revealed preference section, and a stated preference section. Respondents who completed the main survey and consented again were directed to a post-survey on bikeability perceptions.}
  \label{fig:flow}
   \vspace{-0.5cm}
\end{figure}

An overview of the survey design is provided in Figure \ref{fig:flow}. Respondents were sampled using a stratified sampling strategy based on their usual commuting mode, as recorded in the respondent pool data. This approach aimed to balance responses across five commuting modes—Walking, Biking, Car, Public Transit, and Taxi/On-Demand—to ensure sufficient representation for robust mode-specific analysis.

After obtaining explicit consent, respondents were presented with a pre-screener questionnaire. To ensure the survey targeted urban contexts where shared autonomous micro-mobility systems could be viable, we included respondents who fulfilled the following conditions:

\begin{enumerate}
    \item Reside in urban areas with populations exceeding 150,000, excluding small towns and rural areas.
    \item Live within 20 minutes of a public transit stop, excluding suburban areas with limited public transit access.
    \item Had used their reported commuting mode in the past week (or the past month for ride-hailing users).
    \item Know how to ride a bicycle.
\end{enumerate}

Given the lack of real-world data on shared autonomous micro-mobility systems, the survey relied on an SP experiment where respondents chose between hypothetical scenarios. As SP surveys often suffer from hypothetical bias due to the lack of context, we mitigated this by grounding the SP scenarios in recent trips recalled by respondents. Specifically, respondents were asked to describe a recent trip using their commuting mode for which they "could consider using a bicycle or a combination of biking and public transportation". They were also asked to indicate the trip purpose (e.g., work, errands, leisure). Respondents provided origin and destination addresses, which were used—via embedded JavaScript—to query the Google Maps API and retrieve travel time estimates for various modes. These estimates formed the basis for generating the SP scenarios. To protect respondent privacy, trip addresses were used solely for scenario generation and were excluded from the dataset.

\begin{figure}[!htb]
  \centering
  \includegraphics[width=0.7\columnwidth]{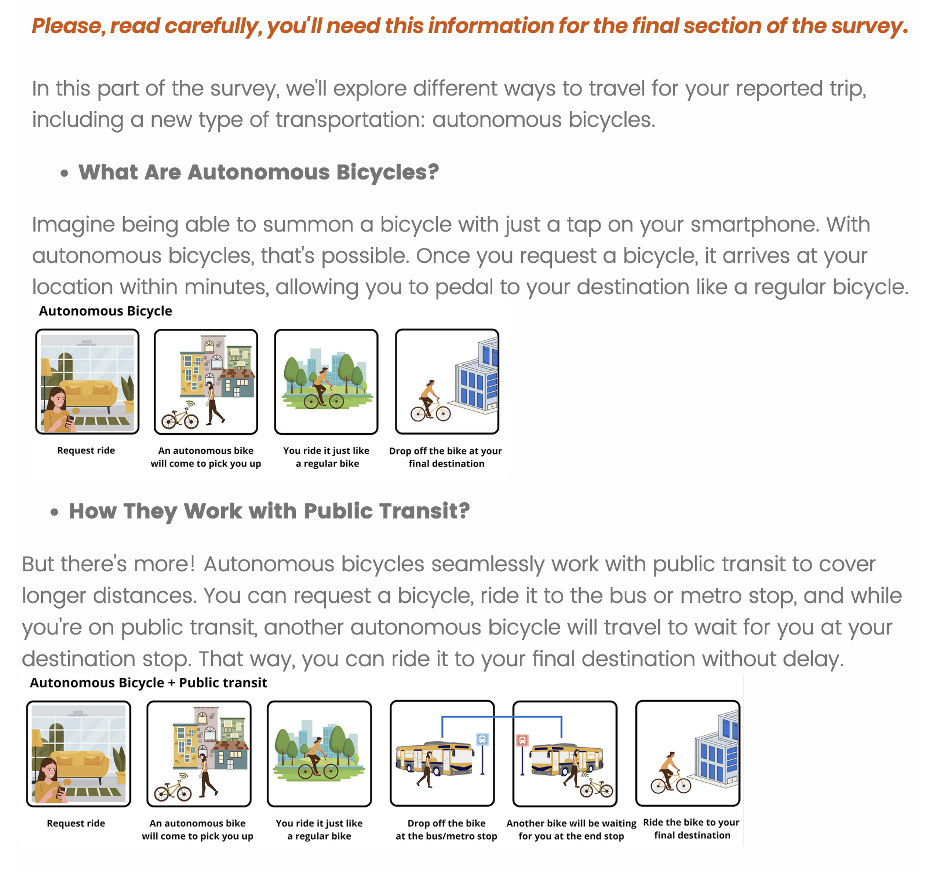}
  \caption{Screenshot of the information shown to respondents when introducing Autonomous Bicycles in the survey. This description was provided prior to the stated preference section to ensure participants had a consistent understanding of the mode.}  \label{fig:introAB}
   \vspace{-0.5cm}
\end{figure}

Following the RP phase, respondents were introduced to the concept of autonomous bicycles through a brief description and accompanying vignettes that explained their use as standalone modes or combined with public transit. The content of this introduction is shown in Figure \ref{fig:introAB}.

\begin{figure}[!htb]
  \centering
  \includegraphics[width=0.6\columnwidth]{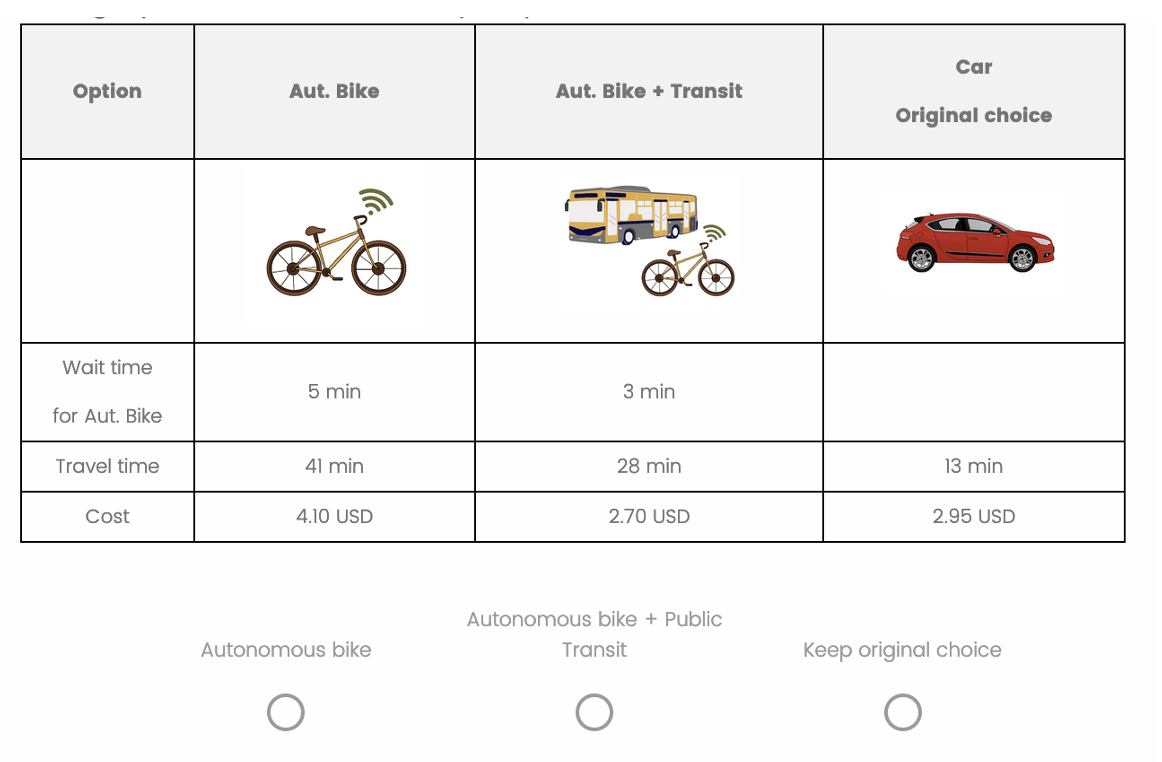}
  \caption{Example of a stated preference (SP) choice task presented to respondents. In each task, participants were asked to choose between keeping their current travel mode or switching to either Autonomous Bicycles or Autonomous Bicycles combined with Public Transit. Each respondent completed six such tasks, with varying cost and wait time attributes for autonomous bicycles and autonomous bicycles with public transit.}
  \label{fig:choiceTask}
   \vspace{-0.5cm}
\end{figure}

Next, respondents completed the SP task, choosing between three alternatives: (1) Autonomous Bicycles, (2) Autonomous Bicycles combined with Public Transit, or (3) Keeping their original mode (e.g., walking, biking, driving, public transit, or taxi/on-demand). Each respondent faced six choice tasks, with varying levels of wait times and costs for autonomous bicycles and autonomous bicycles combined with public transit. The SP scenarios and attribute levels are presented in Table \ref{tab:attributes}, and the format of the choice tasks is shown in Figure \ref{fig:choiceTask}.

Travel times were based on the attributes retrieved during the RP stage for the specific origin and destination of the reported trip. The cost of the currently existing modes was estimated as follows: Car cost was estimated to be \$0.72 USD/mile \citep{bts2024transportation}, public transit cost was estimated to be \$1.5 USD/trip \citep{statista_us_avg_fares}, taxi/on-demand mode was assumed to have a fixed cost of \$1.23 USD, a cost per mile of \$0.97 USD and a cost per minute of \$0.28 USD \citep{lyft_pricing} \footnote{Cost calculated the average cost reported by Lyft \citep{lyft_pricing} for three large cities (New York, Los Angeles, Chicago) and three medium-sized cities (Boston, Portland, Denver).}

For autonomous bicycles, the range for the cost variable was based on existing shared micromobility costs, with the lower end reflecting current average shared bicycle prices (\$0.1 USD/ min) and the higher end being based on the cost of the most costly current e-scooter services (\$1.5 USD/min) \citep{nacto2024shared}. For instance, a 15-minute trip using autonomous bicycles could range from \$1.5 to \$22.5 USD, while a 30-minute trip could range from \$3 to \$45 USD. Costs for autonomous bicycles with public transit were calculated by combining the biking time (first- and last-mile) with the cost of a public transit ride (\$1.5 USD/trip). These ranges were selected to span both typical and upper-bound values observed in current micromobility markets \cite{nacto2024shared}. Additionally, prior work on autonomous micromobility suggests that costs do not necessarily need to exceed current prices in order to be viable or appealing to users \citep{sanchez2020autonomous}.

Wait times ranged from 1 to 15 minutes, based on previous fleet performance estimates for shared autonomous micro-mobility systems \citep{sanchez2022performance}, also aligning with common wait time ranges for current mobility on-demand/ride-hailing services. These ranges are defined with the goal of covering the broadest spectrum of costs that could be expected from shared autonomous micro-mobility, while keeping values realistic enough so that respondents still found the scenarios believable.

\begin{table}[!htb]
\centering
\footnotesize
\begin{tabular}{ll}
\toprule
\textbf{Attribute} & \textbf{Attribute Levels} \\
\midrule
Cost (USD/min)     & 0.1, 0.2, 0.4, 0.7, 1.5 \\
Wait Time (min)    & 1, 3, 5, 7, 10, 15     \\
\bottomrule
\end{tabular}
\caption{Attributes and corresponding levels used in the Stated Preference (SP) scenarios. Each scenario varied in cost and wait time to evaluate respondent sensitivity to these factors when considering Autonomous Bicycle based options.}
\label{tab:attributes}
\end{table}

We employed a full factorial design to generate all possible combinations of cost and wait time attribute levels for the SP scenarios (Table \ref{tab:attributes}). From this design, each respondent was randomly assigned six choice tasks, since the performance of full factorials with a random subset selection has been found to have a comparable performance to d-efficient designs \citep{walker2018d}. The attribute levels for autonomous bicycle-based modes were equal for each scenario because they represent a potential future scenario. The design of the SP scenarios was done programmatically following the procedure outlined in \citep{berke2023drone}, and then imported into the survey platform using embedded HTML, CSS, and JavaScript.

To ensure data quality, the survey included two attention checks: (1) An instructional manipulation check, asking respondents to select a specific number from a set of responses \citep{berke2023drone}. (2) A nonsensical statement,  "At some point in my life, I have had to consume water in some form," with only logically correct answers (Agree/Strongly Agree) accepted \citep{abbey2017attention}. Responses from participants who failed either check were removed from the dataset.

Then, respondents answered eleven attitudinal questions on topics related to biking, public transportation, environmental attitudes, technology, and autonomous bicycles using a 5-point Likert scale. These questions, listed in \ref{app:Indicators}, provided additional insights into respondents' preferences and perceptions and were used to construct the latent variables for the HCM.

Following the main survey, respondents were re-contacted for a follow-up survey on bikeability perceptions. This was necessary because the RP question in the main survey explicitly asked respondents to select a trip for which they could "consider using a bicycle or a combination of biking and public transportation". This framing was chosen to ensure that the scenario was at least potentially relevant to autonomous bicycle use. If respondents had been asked to select any trip, the proportion of autonomous bicycle and autonomous bicycle with public transit choices would have likely been much lower, limiting the model's ability to learn meaningful patterns from the available data. However, this decision also introduced a trip selection bias, since respondents were filtering for potentially bikeable trips.

To address this, the post-survey asked respondents to evaluate a set of eight hypothetical trips of varying lengths and purposes, and indicate whether they considered each trip to be bikeable. These responses were subsequently used to estimate a bikeability binary logit model, as described in Section \ref{sec:tripBias}, allowing us to control for bikeability perception bias in the main model estimation.

The survey received approval from our Institutional Review Board to ensure compliance with ethical research standards for human subjects. 

\subsection{Data Collection}
The survey was built with Qualtrics, a commercial survey-building platform. Responses were collected using Prolific, also a commercial platform, that provides access to a diverse pool of respondents based in the United States (U.S.). The database includes comprehensive demographic information of respondents, such as age, sex, ethnicity, income, family structure, student and employment status, U.S. state of residence, education level, and car ownership. 

As mentioned in the previous section, we used a stratified sampling strategy where respondents were recruited based on their reported commuting mode to work, with the exception of taxi/on-demand ride users. For this last group, respondents were drawn from the general pool and pre-screened by asking whether they had used a taxi or ride-hailing service in the past month. Only those who met this criterion proceeded to the full survey. 

Eligibility criteria required participants to be 18 years or older, and the survey was designed to be accessible on a laptop, tablet, or phone. Participants were compensated at a rate of \$12 USD/hour, which translated to \$1.50 USD for the main survey, which took 5-7 minutes (median of 5.7 minutes), and 0.4 for the 1-minute post-survey. Those who were screened out during the pre-survey phase—lasting an average of 1 minute—were compensated with \$0.20 USD for their time.

Table \ref{tab:respondent_summary} presents a summary including both the total and final number of respondents, as well as the effect of individual pre-screening criteria. Additionally, we excluded responses where the reported travel time for the original mode was less than 1 minute, to filter out potential round trips, and those exceeding 120 minutes, to eliminate inter-city trips or entries with likely address errors. Reported travel distances outside the range of 0.1 to 30 miles were also removed to ensure consistency with realistic inner-city travel patterns.

\begin{table}[!htb]
\centering
\footnotesize
\setlength{\tabcolsep}{6pt} 
\renewcommand{\arraystretch}{1.15}
\caption{Summary of survey respondents and screening criteria. The table reports the total number of initial responses, the number and percentage of respondents excluded based on each pre-screening criterion, the remaining sample after pre-screening, and the final sample size after applying additional filters related to trip duration and distance.}
\label{tab:respondent_summary}
\begin{tabular}{llcc}
\toprule
\textbf{Category} & \textbf{Description} & \textbf{Count} & \textbf{Pct. (\%)} \\
\midrule
\textbf{Total Sample} & All responses collected & 1925 & -- \\
\midrule
\textbf{Pre-Screening } & Not used any mode in the last week & 352 & 18.3 \\
                      & Town/rural area (pop. <150,000) & 544 & 28.3 \\
                      & No transit access (within 20 min) & 525 & 27.3 \\
                      & Cannot ride a bicycle & 118 & 6.1 \\
\midrule
\textbf{Post-Screening } & Sample after pre-screening & 947 & 49.2 \\
\textbf{Final Sample} & After trip-based filtering & 730 & 37.9 \\
\bottomrule
\end{tabular}
\end{table}

\subsection{Sample Bias Correction}

To ensure the generalizability of our results to the broader population, it was necessary to correct for key biases introduced during the survey design and data collection process. Specifically, we addressed three types of bias: (1) trip selection bias, (2) stratified mode sampling bias, and (3) demographic representativeness bias. These corrections relied on a combination of post-survey data and an external benchmark from the National Household Travel Survey (NHTS). We applied an iterative proportional fitting (IPF) procedure to generate sample weights that adjust for these sources of bias.

\subsubsection{Trip Selection Bias} \label{sec:tripBias}

As detailed in Section \ref{sec:survey-design}, the revealed preference (RP) component of our survey asked respondents to report a trip which they would consider bikeable. Therefore, the resulting RP dataset is not representative of all trips taken by the respondent, but rather of a self-selected subset of bikeable trips. 

To correct for this bias, we utilized the data of the follow-up survey (see Section \ref{sec:survey-design}) to estimate a binary discrete choice model of trip likeability. This model was then applied to the full set of trips in the NHTS dataset (see \ref{app:bikeability-model} for model specification and results). This allowed us to filter the NHTS to a set of "bikeable" trips, creating a proper reference population for reweighting. The final weights were therefore constructed to align our sample with NHTS trips deemed plausibly bikeable according to our model. 

As expected, the distribution of modes differs substantially between all NHTS trips and the bikeable subset. For instance, the share of walking trips increases from 7\% to 14\% among bikeable trips, reflecting the intuitive overlap between walkable and bikeable trips. Conversely, car trips drop from 87\% to 78\%, indicating that a significant portion of car trips in the full dataset are not realistically substitutable by bike. 

For population-wide estimations in Section \ref{sec:Results}, we applied predicted mode shift outcomes to the bikeable subset identified in the NHTS, while assuming the original mode distribution remained unchanged for trips classified as non-bikeable.

\subsubsection{Stratified Sampling and Demographic Representativeness Biases}

Beyond trip selection, two additional biases had to be addressed to ensure generalizability: the effect of stratified sampling by mode and the potential demographic imbalances in our survey sample. 

First, the survey was designed with stratified sampling to ensure adequate representation across travel modes. While methodologically useful, this results in mode shares in the sample that do not reflect population-level mode distributions. Second, despite efforts to broadly recruit participants, we could not ensure their demographics perfectly aligned with the broader population.

To correct both biases simultaneously, we used IPF to generate post-stratification weights. These weights were constructed to align the joint distribution of several key variables between our survey sample and a reference dataset of bikeable trips from NHTS, extracted from Section \ref{sec:tripBias}. 

Variables used in the IPF procedure included: trip purpose, mode, sex, age (young/older), race (white), income level (low/high), education level (higher education), student status, full-time employment status, presence of children, hot summer, harsh winter. These variables are defined in Appendix Table \ref{app:variables}

After applying the IPF procedure, the weighted distribution of the survey sample matched the bikeable NHTS benchmark across all reweighted variables. A summary of matched margins after IPF weighting is provided in Appendix Table \ref{tab:ipf_margins}. This confirms that the final sample is representative of the population of bikeable trips along relevant dimensions.

In addition, for analyses that investigate environmental impact differences across urban typologies (Section \ref{sec:env-impacts}), we performed a second reweighting process following the same IPF methodology. In this version, we used bikeable trips from distinct NHTS urban area categories, specifically: 
\emph{Urban areas with 200–499k population}, \emph{500–999k}, \emph{1M+ with rail}, and \emph{1M+ without rail}. These weights enabled us to study how urban contexts influence behavioral responses and impacts in our analysis.

The weights derived through the procedures described in this section were applied in the computation of all aggregate statistics reported in Section \ref{sec:Results}, including mode shares, mode shifts, and environmental impact metrics.

\section{Modeling} \label{sec:modeling}
\subsection{Model Development}

The modeling framework for this study relies on Discrete Choice Modeling (DCM) \citep{mcfadden1972conditional, ben1985discrete,train2009discrete} to estimate mode choice behavior for shared autonomous micro-mobility. This section presents the development of three models: a baseline logit model (Model 1), a mixed logit model with random coefficients and panel structure (Model 2), and a hybrid choice model incorporating a latent variable (Model 3) \citep{ben2002hybrid,abou2024hybrid}. 

While the modeling framework follows standard discrete choice modeling practices, the models themselves are developed specifically for this study. The contribution lies in applying these methods to a novel context: shared autonomous micro-mobility. All three models were developed specifically for this study and estimated using original survey data \ref{sec:surveyandData}. To our knowledge, this is the first application of discrete choice modeling to shared autonomous micro-mobility using attitudinal data and hybrid choice methods.

We followed a stepwise approach, beginning with a simple specification including only cost and time as explanatory variables. We then progressively added trip-specific attributes (e.g., purpose, wait type) and sociodemographic variables (e.g., income, car ownership). Most variables were dummy-coded, and inclusion was guided by statistical significance using likelihood ratio tests at the 95\% confidence level.

\subsubsection{Model 1: Baseline Logit} 
This model includes deterministic utility terms only and assumes homogeneous preferences across individuals. It serves as the reference specification.

\subsubsection{Model 2: Mixed Logit with Panel Effects}  
To account for preference heterogeneity and mode-specific biases, Model 2 extends the baseline specification by introducing random coefficients for travel cost, active travel time, and the alternative-specific constants (ASCs) for autonomous bicycles and autonomous bicycles with public transit. This allows for individual differences in sensitivity to trip attributes and unobserved utility variations across travel alternatives. Additionally, since each respondent completed six choice tasks, we applied a panel structure with an agent effect to capture intra-individual correlation. This accounts for consistency in preferences across repeated SP tasks.

\subsubsection{Model 3: Hybrid Choice Model} 
Lastly, to explicitly account for attitudinal factors influencing adoption of shared autonomous micro-mobility options, Model 3 extends the baseline model by incorporating a latent psychological construct representing negative attitudes toward autonomous bicycles. The latent variable and associated measurement model were newly developed based on factor analysis of attitudinal indicators collected in this study, rather than adopted from existing frameworks or reused from prior hybrid choice studies, ensuring the construct is specific to the context of shared autonomous micro-mobility. In particular, this latent variable was derived from factor analysis of the attitudinal indicators section of the survey (Appendix Table \ref{tab:indicators}). Among three extracted factors: (1) sustainable and active mobility orientation, (2) general technology enthusiasm, and (3) autonomous bicycle perception, we selected the third for model inclusion, as it showed the highest accuracy and was thematically most aligned with the study's focus. Two observed indicators were used to estimate the latent construct: disagreement with the statement (I10) “I think autonomous bicycles would be both convenient and fun” and agreement with (I11) “I would feel uneasy about encountering autonomous bicycles in the streets.” We considered these to represent a latent attitudinal variable representing negative perceptions toward autonomous bicycles.


\subsection{Model Fit}

All models were estimated using maximum likelihood estimation (MLE) in Python Biogeme \citep{bierlaire2023short}.  Estimation results for the three models, including coefficients, standard errors, and model fit statistics, are presented in Table \ref{tab:model_results}. Variable definitions and parameter descriptions are available in Appendix \ref{app:variables} and \ref{app:parameters}. 

Model 3 includes two additional components estimated as part of the hybrid choice framework: (1) a structural equation (Table \ref{tab:sem_results}), which links socio-demographics to the latent variable, and (2) a measurement model, which relates the latent variable to observed attitudinal indicators (Table \ref{tab:measurement_model}). 

\begin{table}[H]
\centering
\scriptsize
\setlength{\tabcolsep}{3pt}
\caption{Model Estimation Results including significance level for attributes ($*p < 0.1$) ($** p < 0.05$) ($*** p < 0.01$)}
\label{tab:model_results}
\begin{tabular}{lcccccc}
\toprule
\textbf{Name} & \textbf{Model 1} & \textbf{Rob. Std Err} & \textbf{Model 2} & \textbf{Rob. Std Err} & \textbf{Model 3} & \textbf{Rob. Std Err} \\
\midrule
ASC\_ab & 0.983*** & 0.152 & 4.12*** & 0.925 & 0.712*** & 0.224 \\
ASC\_ab\_sd & -- & -- & 4.8*** & 0.449 & -- & -- \\
ASC\_abpt & -0.106 & 0.158 & -1.33 & 0.897 & -0.695*** & 0.252 \\
ASC\_abpt\_sd & -- & -- & 7.18*** & 0.705 & -- & -- \\
ASC\_bike & 0.84*** & 0.196 & 3.05** & 1.26 & 1.6*** & 0.259 \\
ASC\_pt & -1.03*** & 0.163 & -2.73** & 1.23 & -0.785*** & 0.235 \\
ASC\_taxi & -0.594*** & 0.156 & -0.161 & 0.828 & 0.591** & 0.23 \\
ASC\_walk & 1.23*** & 0.210 & 3.57*** & 1.36 & 1.87*** & 0.265 \\
B\_activeTime & -4.35*** & 0.241 & -18.9*** & 2.12 & -4.64*** & 0.229 \\
B\_activetime\_sd & -- & -- & 10.8*** & 1.17 & -- & -- \\
B\_carowner & 0.152* & 0.085 & 0.874 & 0.803 & 0.124 & 0.0784 \\
B\_children & 0.555*** & 0.11 & 0.978 & 0.793 & 0.328*** & 0.11 \\
B\_cost & -1.13*** & 0.119 & -7.95*** & 0.731 & -1.17*** & 0.12 \\
B\_cost\_sd & -- & -- & 6.13*** & 0.552 & -- & -- \\
B\_errands & -- & -- & -- & -- & 0.36* & 0.191 \\
B\_fulltime & 0.346*** & 0.0734 & 0.432 & 0.65 & 0.315*** & 0.0724 \\
B\_higher\_ed & 0.259*** & 0.0401 & 0.71** & 0.293 & 0.22*** & 0.0452 \\
B\_highincome & -- & -- & -- & -- & -0.0873 & 0.137 \\
B\_hotsummer & -- & -- & -- & -- & 0.284*** & 0.0614 \\
B\_leisure & 0.191** & 0.0941 & 0.806 & 0.758 & 0.0353 & 0.0965 \\
B\_older & 0.822*** & 0.103 & 3.04*** & 0.646 & 0.878*** & 0.0975 \\
B\_ptshortwait & -- & -- & -- & -- & 0.374*** & 0.0817 \\
B\_time & -1.79*** & 0.231 & -6.52*** & 2.28 & -1.84*** & 0.219 \\
B\_wait & -3.31*** & 0.459 & -9.52*** & 1.19 & -3.75*** & 0.418 \\
B\_white & 0.491*** & 0.117 & 1.56** & 0.771 & 0.477*** & 0.118 \\
B\_work & 0.385*** & 0.0654 & 0.599 & 0.488 & 0.363*** & 0.0629 \\
B\_lv &  - 	& - &- 	&- & 	1.59*** & 0.134 \\

\midrule
\textbf{Model Summary} & & & & & & \\
Estimated Parameters & 18 & -- & 22 & -- & 37 & -- \\
Sample Size & 4446 & -- & 1092 & -- & 4446 & -- \\
Initial Log-Likelihood & -4884.430 & -- & -4035.084 & -- & -22347.340 & -- \\
Final Log-Likelihood & -3513.523 & -- & -2521.121 & -- & -17354.860 & -- \\
Rho-Square-Bar & 0.277 & -- & 0.370 & -- & 0.222 & -- \\
AIC & 7063.046 & -- & 5086.243 & -- & 34783.720 & -- \\
BIC & 7178.241 & -- & 5196.150 & -- & 35020.510 & -- \\
Number of Draws & -- & -- & 2000 & -- & 1000 & -- \\
\bottomrule
\end{tabular}
\end{table}

\begin{table}[H]
\centering
\footnotesize
\caption{Structural Equation Results for Model 3:  Effects of socio-demographics on latent attitudinal variable representing negative perceptions toward autonomous bicycles.}
\label{tab:sem_results}
\begin{tabular}{lcccc}
\toprule
\textbf{Name} & \textbf{Estimate} & \textbf{Std. Err} & \textbf{t-test} & \textbf{p-value} \\
\midrule
coef\_children & -0.494 & 0.0656 & -7.53 & $<0.001$ \\
coef\_higher\_ed & 0.395 & 0.0671 & 5.89 & $<0.001$ \\
coef\_highincome & -0.198 & 0.0859 & -2.31 & 0.0209 \\
coef\_hotsummer & -0.517 & 0.0991 & -5.22 & $<0.001$ \\
coef\_intercept & -1.88 & 0.128 & -14.7 & $<0.001$ \\
coef\_white & 0.271 & 0.0603 & 4.50 & $<0.001$ \\
coef\_woman & 0.31 & 0.0495 & 6.26 & $<0.001$ \\
coef\_young & 0.0372 & 0.0594 & 0.627 & 0.531 \\
\bottomrule
\end{tabular}
\end{table}

\begin{table}[H]
\centering
\footnotesize
\setlength{\tabcolsep}{1pt}
\caption{Measurement Model Results for Model 3: Loadings of indicators on latent attitudinal variable representing negative perceptions toward autonomous bicycles.  For model identification, the loading, intercept, and scale parameters of indicator I10 were fixed.}
\label{tab:measurement_model}
\begin{tabular}{lcccc}
\toprule
\textbf{Indicator} & \textbf{Estimate} & \textbf{Std. Err} & \textbf{t-test} & \textbf{p-value}\\
\midrule
B\_I11  & 0.934 & 0.0527 & 17.7 & $<0.001$ \\
INTER\_I11 & 1.17 & 0.0518 & 22.5 & $<0.001$ \\
SIGMA\_I11 & 1.75 & 0.15 & 11.7 & $<0.001$  \\
\bottomrule
\end{tabular}
\end{table}

\subsection{Model Selection}

Since Model 3 is estimated as a hybrid choice model, its log-likelihood reflects the joint likelihood of both observed choices and attitudinal indicator responses. As a result, metrics such as log-likelihood, $\bar{\rho}^2$,  AIC, and BIC are not directly comparable across Models 1–3.

To assess model performance in a way that supported consistent comparison and the prediction goals of this study, we used two out-of-sample metrics: (1) average predictive accuracy across all modes, and (2) mean absolute difference between predicted and actual mode shares. Both metrics were computed using five-fold cross-validation. 

\begin{figure}[H]
\centering
\includegraphics[width=\textwidth]{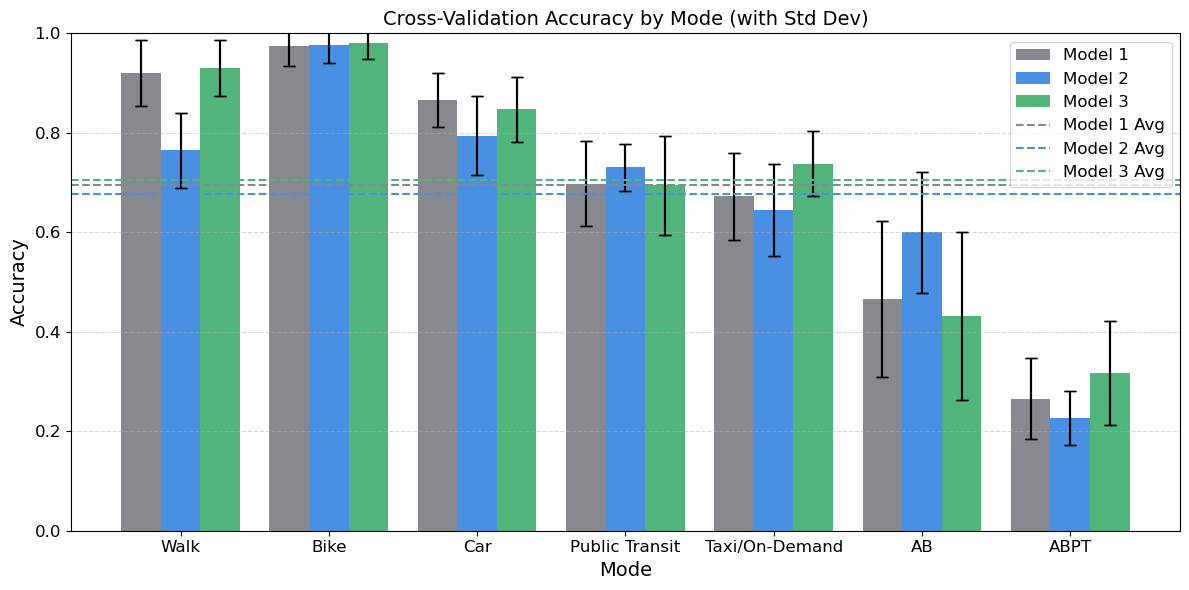}
\caption{Average predictive accuracy across all modes for Models 1–3, based on five-fold cross-validation.}
\label{fig:cv_accuracy}
\end{figure}

\begin{figure}[H]
\centering
\includegraphics[width=\textwidth]{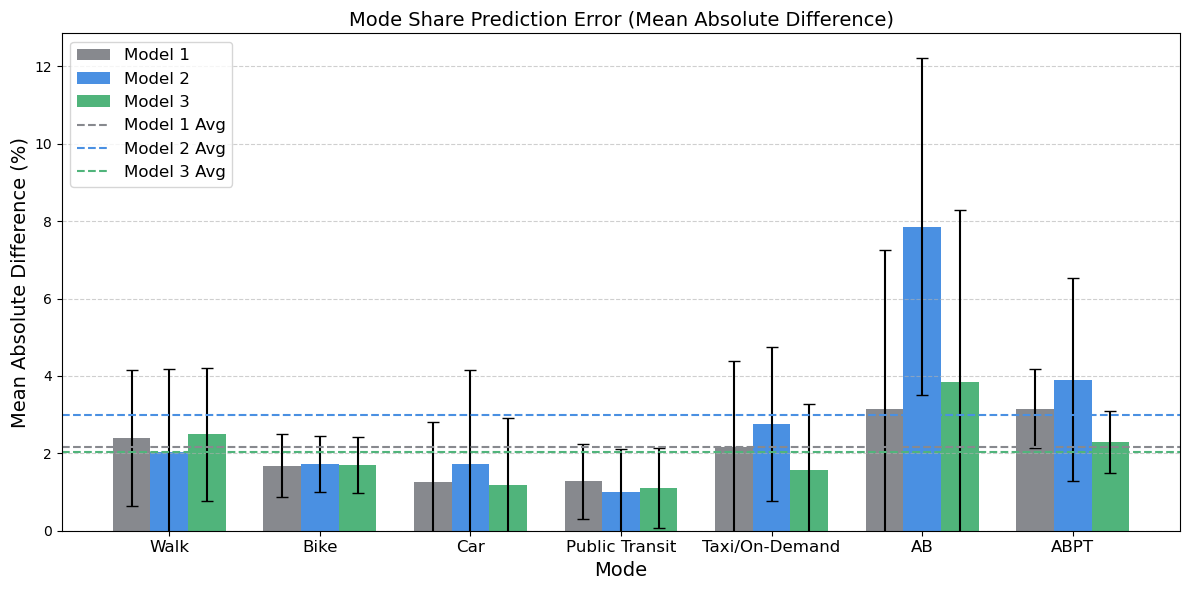}
\caption{Mean absolute difference (\%) in predicted mode shares for Models 1–3, based on five-fold cross-validation.}
\label{fig:mode_error}
\end{figure}

As shown in Figures \ref{fig:cv_accuracy} and \ref{fig:mode_error}, Model 3 consistently outperformed the other two models. It achieved the highest average accuracy across modes and the lowest overall error in predicted mode shares. While the improvements are modest in magnitude, they are consistent across modes and reflect better generalization rather than overfitting, as they are based on out-of-sample performance using five-fold cross-validation. Importantly, the hybrid choice model also provides behavioral insight by incorporating a latent attitudinal construct, which cannot be captured by Models 1 or 2. Given this combination of improved predictive performance and enhanced interpretability,  Model 3 was selected for scenario simulations and all subsequent results presented in Section \ref{sec:Results}. Full numerical results are provided in \ref{app:cv-tables}. 

\subsection{Final Model Specification}

The final specification corresponds to Model 3, which combines a discrete choice component with a latent variable capturing negative attitudes toward autonomous bicycles. The observable utility functions $V_i$ are specified as:

\[
V_i = ASC_i + LV_{\text{effect}} + \sum_x B_x \cdot x_i
\]

where $V_i$ is the utility of alternative $i$, $ASC_i$ is the alternative-specific constant, $B_x$ is the coefficient associated with explanatory variable $x$, and $x_i$ is the value of $x$ for alternative $i$.

In addition to standard observable attributes, the HCM includes an attitudinal latent variable that influences preferences for autonomous bicycle-based alternatives. This latent variable, denoted $LV_{\text{AB}}$, represents negative perceptions toward autonomous bicycles and is included in the utility functions of autonomous bicycles and autonomous bicycles with public transit.

\paragraph{Structural Equation}  
The latent construct $LV_{\text{AB}}$ is modeled as a linear function of socio-demographic variables and a normally distributed random term:

\[
LV_{\text{AB}} = \text{coef}_{\text{intercept}} + \sum_j \text{coef}_j \cdot z_j + \sigma_s \cdot \omega
\]

where $z_j$ are individual characteristics, $\sigma_s$ is the standard deviation of the latent variable, and $\omega \sim \mathcal{N}(0, 1)$ is a simulated standard normal draw.

To incorporate $LV_{\text{AB}}$ into the utility functions, we apply a scaled hyperbolic tangent transformation:

\[
LV_{\text{effect}} = - B_{\text{lv}} \cdot \tanh(LV_{\text{AB}})
\]

This nonlinear transformation ensured a bounded influence of attitudes on mode utility and contributed to model convergence and stability.

\paragraph{Measurement Model}  

I10 was used as the reference indicator, with its intercept, loading, and scale fixed for identification. I11's parameters were estimated using an ordered probit model. 

Let \( y_i^* \) denote the unobserved continuous response for each indicator, and \( y_i \in \{1, 2, 3, 4, 5\} \) the observed Likert response. Then:

\[
y_i^* = \alpha + \lambda \cdot LV_{\text{AB}} + \varepsilon_i, \quad \varepsilon_i \sim \mathcal{N}(0, 1)
\]

\[
y_i = k \quad \text{if} \quad \tau_{k-1} < y_i^* \leq \tau_k \quad \text{for } k = 1,\dots,5
\]

The thresholds were symmetrically parameterized using two parameters \( \delta_1 \) and \( \delta_2 \):

\[
\tau_1 = -\delta_1 - \delta_2,\quad
\tau_2 = -\delta_1,\quad
\tau_3 = \delta_1,\quad
\tau_4 = \delta_1 + \delta_2
\]

with \( \tau_0 = -\infty \) and \( \tau_5 = \infty \).

\paragraph{Utility Functions}  
The utility functions $V_0$ to $V_6$ correspond to the seven travel alternatives: Walk, Bike, Car, Public Transit, Taxi/On-Demand, Autonomous Bikes, and Autonomous Bikes in combination with Public Transit. They are defined as:

\begin{align*}
V_0 &= ASC_{\text{walk}} + B_{\text{activetime}} \cdot \mathit{WalkTime} + B_{\text{white}} \cdot \mathit{white} + B_{\text{higher\_ed}} \cdot \mathit{higher\_ed} \\
&\quad - B_{\text{children}} \cdot \mathit{children} - B_{\text{hotsummer}} \cdot \mathit{hot\_summer} \\[1ex]
V_1 &= ASC_{\text{bike}} + B_{\text{activetime}} \cdot \mathit{BikeTime} - B_{\text{carowner}} \cdot \mathit{car\_owner} \\
&\quad - B_{\text{highincome}} \cdot \mathit{high\_income} 
- B_{\text{fulltime}} \cdot \mathit{full\_time} + B_{\text{higher\_ed}} \cdot \mathit{higher\_ed} \\
&\quad - B_{\text{leisure}} \cdot \mathit{leisure\_trip} + B_{\text{work}} \cdot \mathit{work\_trip} \\[1ex]
V_2 &= ASC_{\text{car}} + B_{\text{cost}} \cdot \mathit{CarCost} + B_{\text{time}} \cdot \mathit{CarTime} + B_{\text{carowner}} \cdot \mathit{car\_owner} \\
&\quad - B_{\text{white}} \cdot \mathit{white} + B_{\text{older}} \cdot \mathit{older} - B_{\text{leisure}} \cdot \mathit{leisure\_trip} \\
&\quad - B_{\text{work}} \cdot \mathit{work\_trip} + B_{\text{errands}} \cdot \mathit{errands\_trip} \\[1ex]
V_3 &= ASC_{\text{pt}} + B_{\text{cost}} \cdot \mathit{PTCost} + B_{\text{time}} \cdot \mathit{TransitTime} + B_{\text{fulltime}} \cdot \mathit{full\_time} \\
&\quad + B_{\text{higher\_ed}} \cdot \mathit{higher\_ed} - B_{\text{children}} \cdot \mathit{children} - B_{\text{leisure}} \cdot \mathit{leisure\_trip} \\
&\quad + B_{\text{work}} \cdot \mathit{work\_trip} - B_{\text{hotsummer}} \cdot \mathit{hot\_summer} + B_{\text{ptshortwait}} \cdot \mathit{pt\_shortWait} \\[1ex]
V_4 &= ASC_{\text{taxi}} + B_{\text{cost}} \cdot \mathit{TaxiCost} + B_{\text{time}} \cdot \mathit{CarTime} + B_{\text{wait}} \cdot \mathit{TaxiWaitTime} \\
&\quad + B_{\text{highincome}} \cdot \mathit{high\_income} + B_{\text{leisure}} \cdot \mathit{leisure\_trip} + B_{\text{hotsummer}} \cdot \mathit{hot\_summer} \\[1ex]
V_5 &= ASC_{\text{ab}} + LV_{\text{effect}} + B_{\text{cost}} \cdot \mathit{ABCost} + B_{\text{activetime}} \cdot \mathit{BikeTime} \\
&\quad + B_{\text{wait}} \cdot \mathit{ABWaitTime} - B_{\text{fulltime}} \cdot \mathit{full\_time} - B_{\text{older}} \cdot \mathit{older} \\
&\quad - B_{\text{higher\_ed}} \cdot \mathit{higher\_ed} - B_{\text{work}} \cdot \mathit{work\_trip} \\[1ex]
V_6 &= ASC_{\text{abpt}} + LV_{\text{effect}} + B_{\text{cost}} \cdot \mathit{ABPTCost} + B_{\text{activetime}} \cdot \mathit{ABPTBikeTime} \\
&\quad + B_{\text{time}} \cdot (\mathit{ABPTTotalTime} - \mathit{ABPTBikeTime}) + B_{\text{wait}} \cdot \mathit{ABWaitTime} \\
&\quad - B_{\text{carowner}} \cdot \mathit{car\_owner} - B_{\text{higher\_ed}} \cdot \mathit{higher\_ed} \\
&\quad + B_{\text{hotsummer}} \cdot \mathit{hot\_summer} + B_{\text{ptshortwait}} \cdot \mathit{pt\_shortWait}
\end{align*}

The utilities are input into a multinomial logit choice model with the appropriate availability structure. 

\section{Results}\label{sec:Results}

In this section, we present the key findings from the final model, Model 3.We begin by interpreting the behavioral drivers of mode choice, then assess mode shifts under various scenarios, and finally quantify the resulting environmental impacts.

It should be kept in mind that this study is focused on the United States context, and the survey was administered only to U.S.-based respondents. As such, findings may not generalize to other countries with different transportation systems, regulatory environments, or cultural attitudes toward new mobility technologies.

\subsection{Behavioral Insights from Final Model}

The estimated coefficients from Model 3 are summarized in Table \ref{tab:model_results}, Table \ref{tab:sem_results}, and Table \ref{tab:measurement_model}. Estimated parameters align with theoretical expectations in both sign and magnitude. As expected, cost, time, and wait variables consistently enter with negative signs, confirming their disutility in mode choice.

The estimated value of time (VOT) metrics indicate meaningful distinctions across components of travel. The value of active time (walking and biking) is estimated at \$39.6/h, followed by wait time at \$32.1/h, and general travel time at \$15.7/h. These results suggest that respondents perceive active and waiting time as significantly more burdensome than overall trip time. The relatively high disutility associated with active time may reflect physical effort, discomfort due to environmental factors, or safety concerns. Similarly, the disutility of wait time may stem from a reduced sense of control and uncertainty during the waiting period. 

Trip characteristics also influence mode choice. Work trips are associated with a lower likelihood of choosing autonomous bicycles, a preference that is also present among full-time employed individuals. These segments may place greater value on reliability and time, and be less inclined to adopt a new mode whose performance characteristics they are less familiar with. Errand trips are positively associated with car use, which aligns with the greater flexibility and carrying capacity often required for such purposes.

Sociodemographic variables also play an important role. Older individuals are significantly less likely to choose autonomous bicycles, potentially reflecting lower willingness to engage in active or novel forms of mobility. Respondents with higher levels of education also exhibit reduced preference for autonomous bicycles both alone or with public transit, which may indicate more critical perceptions associated with new technologies.

Car ownership is negatively associated with the choice of autonomous bicycles with public transit, but not autonomous bicycles alone. This suggests that individuals with private vehicle access may be more reluctant to include public transport in their travel chain, possibly due to concerns over transfers, delays, or perceived inconvenience. In contrast, shorter public transit wait times are positively associated with the use of autonomous bicycles in combination with public transit, underscoring the importance of minimizing transfer-related burdens in multi-modal trips.

The latent variable representing negative attitudes toward autonomous bicycles enters significantly and negatively in the utility functions of both autonomous bicycles and autonomous bicycles with public transit, indicating that attitudinal factors play a significant role in mode choice. As shown in Table \ref{tab:sem_results}, negative attitudes towards autonomous bicycles are more likely to happen among women, white respondents, and those with higher education, while the presence of children, higher income, and residence in hot climates are less likely to have this negative attitude. Most of these relationships are consistent with broader patterns in the adoption of biking and new technologies.

Importantly, the inclusion of this latent construct not only improves model fit but also underscores the influence of psychological factors in shaping behavioral responses. Mode choices are influenced not only by objective attributes such as travel time or cost, but also by individuals’ subjective perceptions and attitudes. 

\subsection{Sociodemographic Characteristics of Autonomous Bicycle Adopters}

\begin{figure}[H]
\centering
\includegraphics[width=0.9\textwidth]{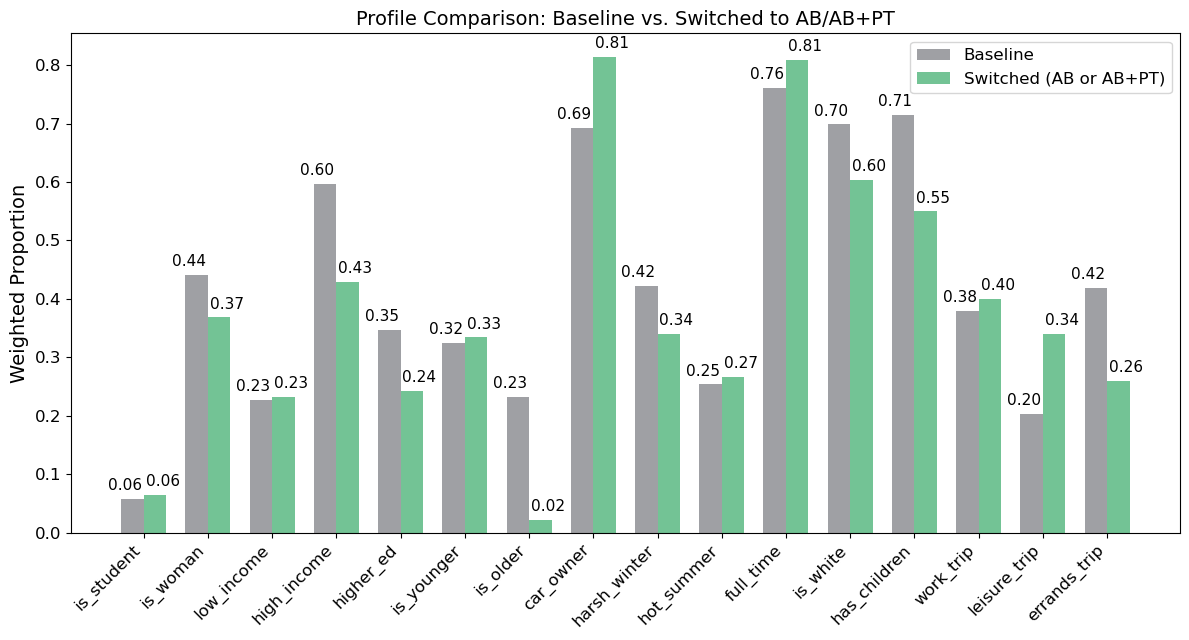}
\caption{Weighted sociodemographic and trip attribute proportions for the full sample (Baseline) and for respondents who selected Autonomous Bicycle (AB) or Autonomous Bicycles in combination with Public Transit (ABPT) across scenarios.}
\label{fig:profile_comparison}
\end{figure}

To complement the model-based behavioral insights, we compare the demographic and trip profiles of the overall sample to those who selected autonomous bicycles either alone or in combination to public transit across all wait time and cost scenarios. Figure \ref{fig:profile_comparison} presents weighted proportions for selected attributes across the full respondent pool and the subset of respondents who chose an autonomous bicycle mode.

This descriptive comparison reveals several notable differences in the characteristics of respondents who selected autonomous bicycle options. Compared to the full sample, these adopters are underrepresented among women, individuals with high income or higher education, older adults, respondents living in states with harsh winters, white individuals, those with children, and those traveling for errands. Conversely, they are overrepresented among car owners, full-time workers, and individuals taking leisure trips. While these patterns offer insight into adoption tendencies, they do not imply causal relationships and should be interpreted as indicative of associations rather than determinants.

It is important to note that observed sociodemographic patterns diverge in important ways from existing trends in shared micromobility use. Prior studies have found that traditional bike-sharing systems in the U.S. tend to be disproportionately used by higher-income, white, highly educated men \citep{fishman2016bikeshare, berke2024access}. In contrast, this analysis suggests that autonomous bicycle adoption may exhibit a more balanced demographic profile—particularly with respect to income, race, and education—when systems are assumed to be uniformly available across urban areas. This points to a potential for broader appeal under conditions of equitable access.

Nonetheless, other forms of disparity remain. Adoption is still skewed by age and gender, with lower participation observed among older adults, women, and individuals with children. These patterns are consistent with prior research identifying safety concerns, perceived physical demands, and caregiving responsibilities as key barriers to cycling and new mobility adoption within these groups \citep{graystone2022gendered, cerin2017neighbourhood}. Some of these challenges could be addressed through targeted design improvements aimed at enhancing comfort and safety, for example, incorporating weather protection, greater vehicle stability, or cargo-capable designs \citep{lin2021affordable}. Such features may improve usability for individuals who need to transport goods or travel with children, thereby making the system more suitable for errands and utility-oriented trips. It is also important to consider users with disabilities, who were not explicitly captured in this analysis but may face additional barriers depending on the type of impairment and the vehicle's accessibility features.

Overall, these findings highlight the importance of designing autonomous mobility systems that are not only technically feasible but also inclusive by design. A diversified fleet that accommodates a range of user needs and trip purposes can help reduce demographic disparities in adoption. In addition, equitable deployment strategies and widespread safe infrastructure will be essential to avoid reproducing the access and usage biases observed in earlier mobility services.

\subsection{Predicted Mode Shares and Mode Shifts}\label{sec:mode-shift}

This section examines the predicted adoption of autonomous bicycle-based modes under varying cost and wait time conditions. Simulations are based on the attribute levels used in the stated preference survey design (Table~\ref{tab:attributes}).

\begin{figure}[H]
\centering
\includegraphics[width=\textwidth]{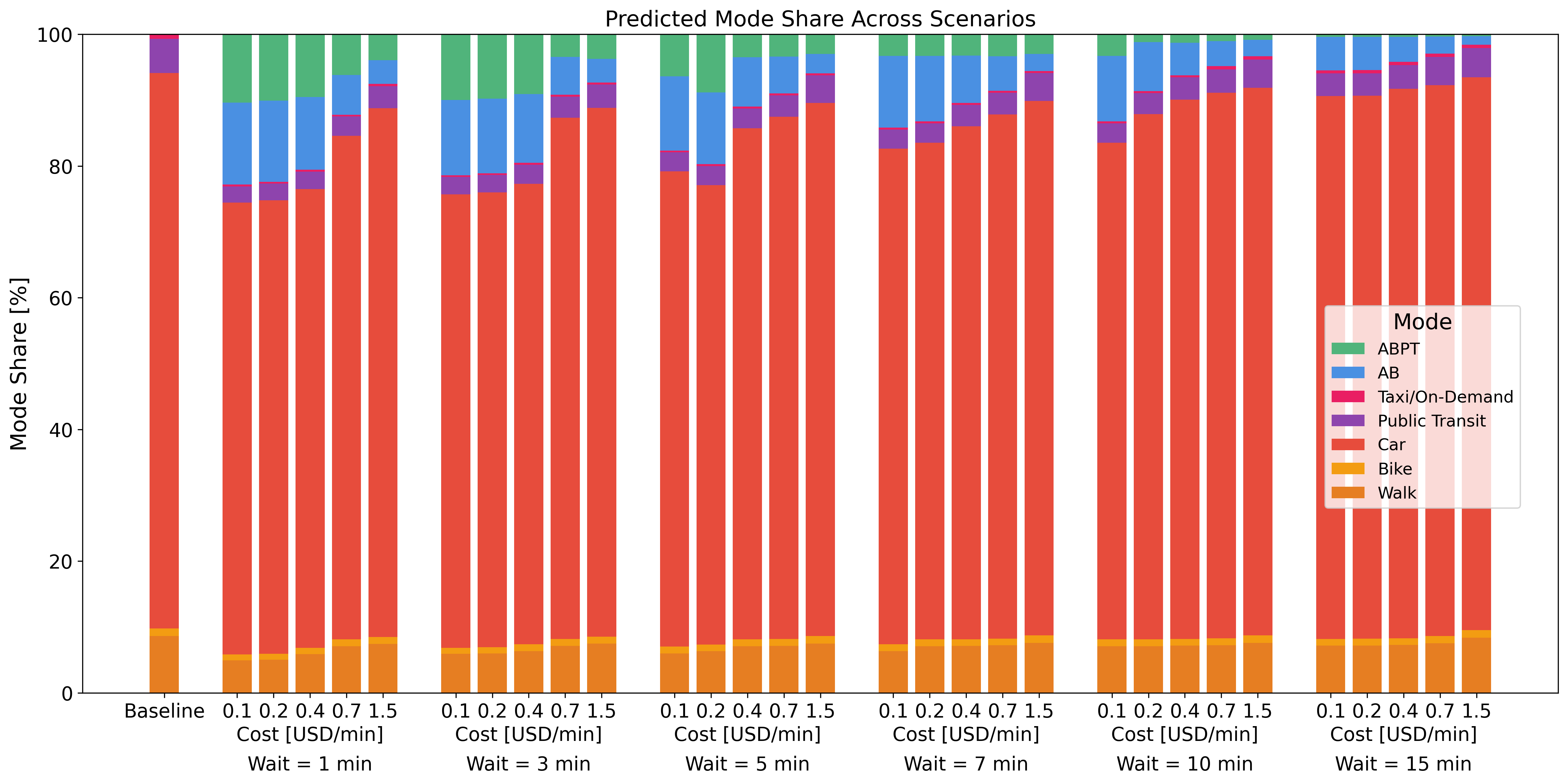}
\caption{Predicted total mode shares across cost and wait time scenarios for Autonomous Bicycles (AB) and Autonomous Bicycles combined with Public Transit (ABPT), compared to baseline mode shares. Each group of bars corresponds to a fixed wait time, with increasing per-minute cost from left to right within each group. This includes both bikeable and non-bikeable trips.}
\label{fig:mode_shares}
\end{figure}

\begin{figure}[H]
\centering
\includegraphics[width=\linewidth]{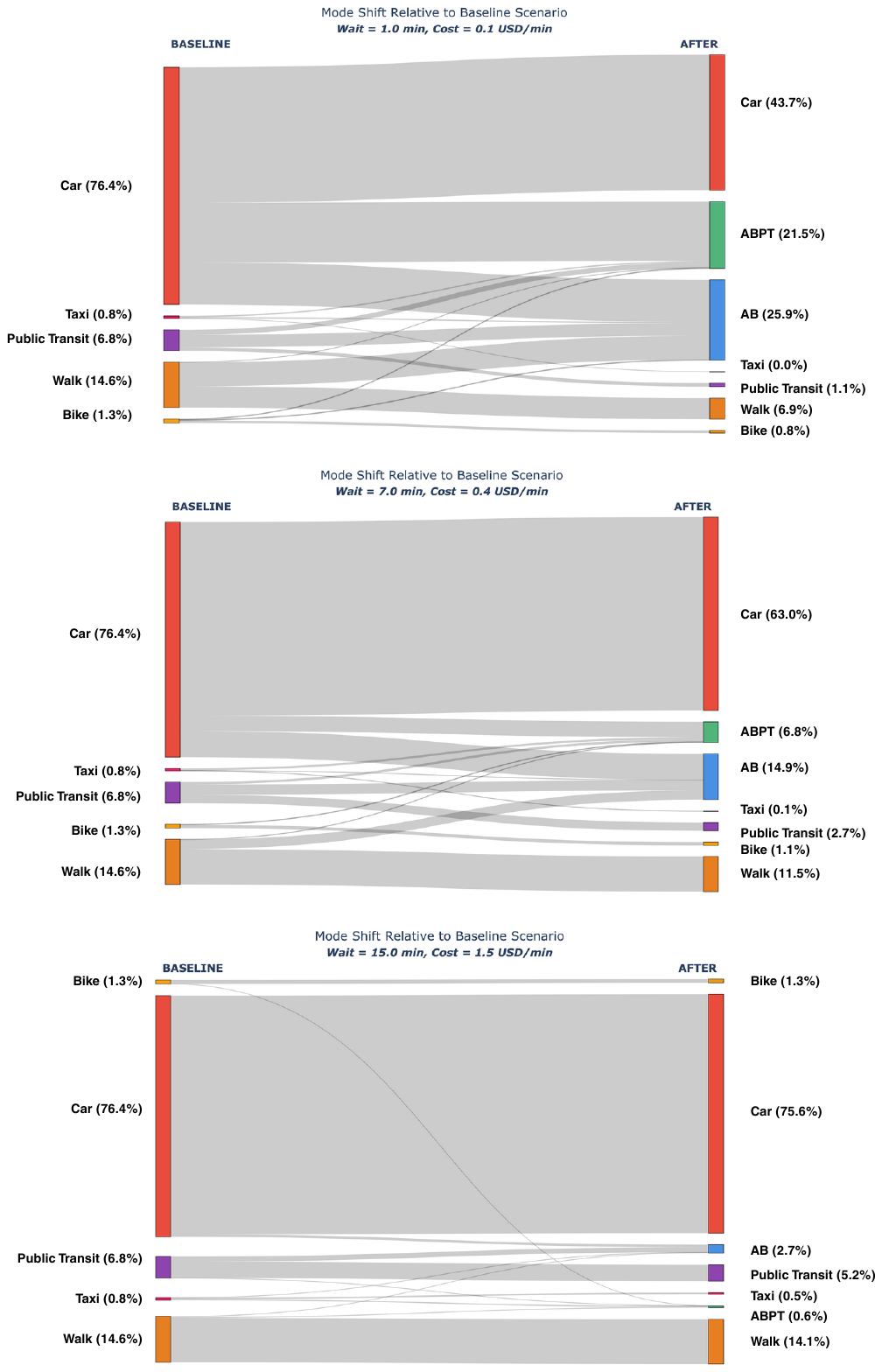}
\caption{Mode shift within \textit{bikeable} trips from baseline to scenarios with autonomous bicycles (AB) and AB combined with public transit (ABPT) under varying wait time and cost conditions: (a) Low cost, low wait (Wait = 1 min, Cost = \$0.1 USD/min), (b) Medium cost, medium wait (Wait = 7 min, Cost =\$ 0.4 USD/min), (c) High cost, high wait (Wait = 15 min, Cost = \$1.5 USD/min).}
\label{fig:sankeys}
\end{figure}

Predicted \textit{total} mode shares across the modeled scenarios are shown in Figure \ref{fig:mode_shares}. As expected, increasing either wait time or cost leads to a substantial reduction in the overall penetration of autonomous bicycle-based modes. Figure \ref{fig:sankeys} illustrates the origin of trips shifting to autonomous bicycles or autonomous bicycles with public transit for the \textit{subset of bikeable} trips in three representative scenarios: the most favorable (lowest wait and cost), the most constrained (highest wait and cost), and an intermediate case. 

Under the most favorable conditions—lowest cost (\$0.1 USD/min) and shortest wait (1 minute)—autonomous bicycles and autonomous bicycles with public transit together account for 22.8\% of total demand. Since results are reweighted using NHTS data to ensure generalizability, this share reflects the proportion of all urban trips in U.S. cities with more than 200,000 inhabitants and access to public transit. In this scenario, car mode share drops from 84.4\% to 68.6\%, representing a very significant shift away from private vehicle use. As shown in Figure \ref{fig:sankeys}, autonomous bicycles draw users from car, walking, and public transit, while autonomous bicycles in combination with public transit gaining users primarily coming from car users. Although autonomous bicycles achieve high uptake, the substantial substitution from walking and transit may limit their net environmental benefits, as will be discussed in Section \ref{sec:env-impacts}. In contrast, when combined with public transit, it appears more targeted in displacing car trips, suggesting greater potential environmental impact reductions under these conditions.

As wait time and cost increase, adoption declines noticeably. In the medium cost and wait scenario (\$0.4 USD/min and 7-minute wait), the combined share of autonomous bicycles alone and in combination with public transit falls to 11\%. As shown in Figure \ref{fig:sankeys}, the combination with public transit continues to attract a significant share of former car users' bikeable trips, indicating that moderate delays remain acceptable when prices are competitive with driving. For autonomous bicycles alone, the share of trips originating from walking decreases, while those from transit increase slightly. This likely reflects the growing influence of wait time in short trips, where 7 minutes of waiting may constitute a significant share of total travel time.

In the highest cost, highest wait scenario (\$1.5 USD/min and 15-minute wait), autonomous bicycles alone and with public transit account for only 1.6\% of the total demand. As illustrated in Figure \ref{fig:sankeys}, the system becomes significantly less attractive across all modes, with most users reverting to their original choices. Autonomous bicycles draw a small number of transit users, while the combination with public transit sees minimal uptake, primarily from taxi and on-demand services. The need to wait twice—once for the autonomous bicycle and again for public transit—likely makes this combined mode particularly unappealing in this context, especially for car users.

More broadly, as seen in Appendix Figure \ref{fig:ab_shift}, autonomous bicycle adoption frequently comes from users of already-sustainable modes, at rates that depend on the specific scenario. The shift from conventional biking is limited across all cases, which aligns with expectations: autonomous bicycles are a more expensive alternative that requires waiting, whereas personal bicycles have no cost per ride and have no wait time. For autonomous bicycles combined with public transit, the majority of mode switching comes from car users across all scenarios with wait times below 10 minutes (Appendix Figure \ref{fig:abpt_shift}). This is likely due to a higher suitability for longer trips, where a full autonomous bicycle ride may be impractical, but combining autonomous bicycles with transit offers a viable alternative. Taxi and on-demand users also shift to autonomous bicycles combined with public transit at a steady rate, even under higher wait conditions and cost, likely because their original mode already involves waiting and a higher cost. In contrast, walk trips rarely shift to autonomous bicycles combined with public transit, which may be explained by the fact that walking is typically used for shorter distances, where the added time and coordination of a multimodal trip offer little benefit.

\subsection{Environmental impacts}\label{sec:env-impacts}

To assess the environmental implications of autonomous bicycle adoption, we simulate a range of plausible background conditions and technology assumptions across different wait and cost scenarios. These vary in terms of (i) the emissions profile of non-autonomous bicycle based modes (i.e., representing the surrounding vehicle fleet), and (ii) key lifecycle characteristics of autonomous bicycle systems.

\begin{table}[htbp]
\centering
\footnotesize
\caption{Background fleet emissions assumptions for non-autonomous bicycle based modes. Values are in grams of CO$_2$e per passenger-kilometer (pkm), adapted from \cite{sanchez2022can}.}
\label{tab:emissions_scenarios}
\begin{tabular}{lcccc}
\toprule
\textbf{Scenario} & \textbf{Car} & \textbf{Taxi/On-Demand} & \textbf{Bike} & \textbf{Walk} \\
\midrule
High Emissions & 162 & 91 & 24 & 0 \\
Low Emissions  & 108 & 52 & 0  & 0 \\
Mixed          & 135 & 72 & 12 & 0 \\
\bottomrule
\end{tabular}
\end{table}

Table \ref{tab:emissions_scenarios} summarizes the assumed emissions factors, derived from \cite{sanchez2022can}. Walking emissions are assumed to be zero across all scenarios. For biking, we use regular bicycles in the low emissions scenario and electric bicycles in the high emissions scenario. For private cars, we assume internal combustion engine (ICE) cars in the high emissions scenario and battery electric vehicles (BEVs) in the low emissions scenario. For taxi/on-demand services, ICE taxis are used in the high scenario, and BEV ride-hailing is used in the low emissions scenario. Lastly, for public transit, we assume ICE buses in the high emissions scenario and metro systems in the low emissions scenario. The mixed scenario averages the values across these two extremes.

We also account for variability in autonomous bicycle lifecycle emissions across wait time scenarios by linking per-kilometer emissions to average vehicle utilization (i.e., trips/bike/day). Following the results of \cite{sanchez2022performance}, we estimate the average number of trips per vehicle per day at each wait time level and derive corresponding equivalent emissions per passenger-kilometer traveled using the methodology in \cite{sanchez2022can}. This approach captures a key operational tradeoff: achieving shorter average wait times—such as 1 minute—requires deploying a significantly larger fleet, which in turn lowers the average utilization per vehicle and increases emissions per passenger kilometer traveled. Conversely, longer wait times allow for higher vehicle utilization and result in lower per-kilometer emissions.

Also following the methodology and tool in \cite{sanchez2022can}, in addition to baseline emissions for autonomous bicycles, we consider three variations key lifecycle characteristics: (i) a long lifespan scenario, in which vehicle lifespans are extended to five years (compared to three years in the baseline); (ii) a short lifespan scenario, assuming only one year of service life; and (iii) a high infrastructure scenario, where emissions from supporting infrastructure are doubled relative to the baseline. These estimates also incorporate the relationship between wait time and vehicle utilization. Table~\ref{tab:ab_scenarios} summarizes the resulting emissions across scenarios. These estimates assume a mixed background fleet as the baseline for comparative impact analysis.

\begin{table}[htbp]
\centering
\footnotesize
\setlength{\tabcolsep}{3.5pt} 
\caption{Emissions per kilometer for autonomous bicycles (AB) under different lifecycle assumptions and wait times. Values are in grams of CO$_2$e per passenger-kilometer (pkm), based on \cite{sanchez2022can}}
\label{tab:ab_scenarios}
\begin{tabular}{llcccccc}
\toprule
\textbf{Scenario} & \textbf{AB Variant} & \textbf{1 min} & \textbf{3 min} & \textbf{5 min} & \textbf{7 min} & \textbf{10 min} & \textbf{15 min} \\
\midrule
\multirow{4}{*}{Mixed Fleet} 
 & Baseline         & 83.5  & 57.5  & 45.2  & 42.5  & 40.0  & 38.1 \\
 & Long Lifespan    & 63.9  & 48.3  & 40.9  & 39.3  & 37.8  & 36.7 \\
 & Short Lifespan   & 181.8 & 103.7 & 66.9  & 58.6  & 51.2  & 45.6 \\
 & High Infrastructure & 104.6 & 78.6 & 66.4  & 63.6  & 61.2  & 59.3 \\
\bottomrule
\end{tabular}
\end{table}

\begin{sidewaysfigure}
\centering
\includegraphics[width=\textwidth]{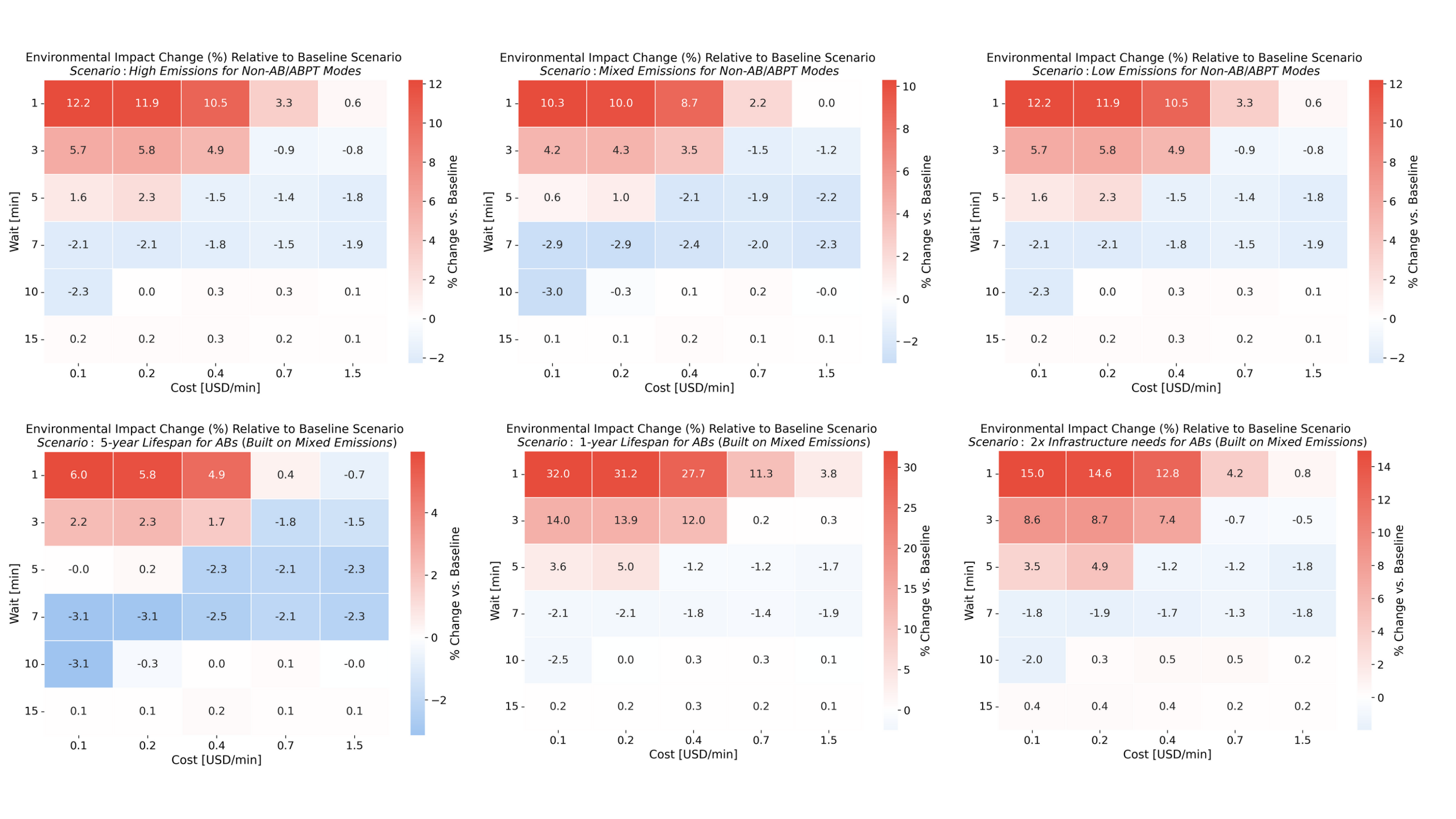}
\caption{Relative change in environmental impact (\%) compared to the baseline scenario across cost and wait time scenarios. Each subplot reflects a different emissions or lifecycle assumption: (top row, left to right) High Emissions, Mixed Emissions, and Low Emissions scenarios for non-autonomous bicycle based modes; (bottom row, left to right) Long Lifespan, Short Lifespan, and High Infrastructure assumptions for autonomous bicycles. Negative values (in blue) indicate environmental improvements.}
\label{fig:env_heatmaps}
\end{sidewaysfigure}

Figure \ref{fig:env_heatmaps} presents the percentage change in environmental impact relative to the baseline scenario with no autonomous bicycles. Each subplot corresponds to a distinct background scenario emissions or autonomous bicycle system configuration, displaying results across all combinations of wait time and cost. The largest \textit{increases in emissions} occur under low-wait and low-cost conditions. As seen in Figure \ref{fig:sankeys} in such scenario, autonomous bicycle adoption is widespread but draws a significant share of users from walking and public transit. Additionally, it has high fleet size requirements and consequently limited per-vehicle utilization. In particular, the worst-case increase—over 32\%—is observed in the 1-year lifespan scenario at 1-minute wait and \$0.1 USD/min. Even when assuming a 5-year lifespan, this low-cost, low-wait setting produces an 6\% emissions increase.

By contrast, \textit{emissions reductions} are observed across multiple scenarios in the mid-range of the design space. In particular, wait times of 7–10 minutes paired with moderate or low costs consistently yield environmental benefits, as well as slightly lower times (5 minutes) at a higher cost. The largest reductions—around 3.6\%—occur when the wait time is 7 minutes, the cost is \$0.1 USD/min, and the rest of the background vehicles are ICE. In addition, a long lifespan also benefits the reduction in emissions. These benefits reflect a combination of targeted mode replacement and efficient vehicle use, as longer wait times reduce fleet size and increase the number of daily trips per vehicle.

It is interesting to note that the increase/reductions pattern across wait and cost combinations remains robust across system assumptions: low wait and low cost scenarios risk environmental increases, while moderate wait and/or cost can act as natural filters, guiding autonomous bicycle adoption away from low-emission trips and toward higher-impact modes like cars. Intuitively, the overall magnitude of the effect decreases as the mode share decreases, which is shown by the small changes in high-cost high-wait scenarios.

These results suggest a fundamental tradeoff: while low wait times and prices may maximize adoption and, therefore, market penetration, they can also yield net environmental impact increases. This tension underscores the importance of aligning service design with broader sustainability goals.

\clearpage
\clearpage

\subsubsection{Variation by Urban Typology}

To investigate how environmental impacts may vary across urban typologies, we study the different urban areas described in Section \ref{sec:tripBias}. This analysis draws on four NHTS-defined urban area categories: \emph{200–499k population}, \emph{500–999k}, \emph{1M+ with rail}, and \emph{1M+ without rail}. These strata allow us to isolate how these urban typologies influence both baseline mode shares and the resulting environmental effects of autonomous bicycle deployment.

\begin{figure}[htbp]
\centering
\includegraphics[width=\textwidth]{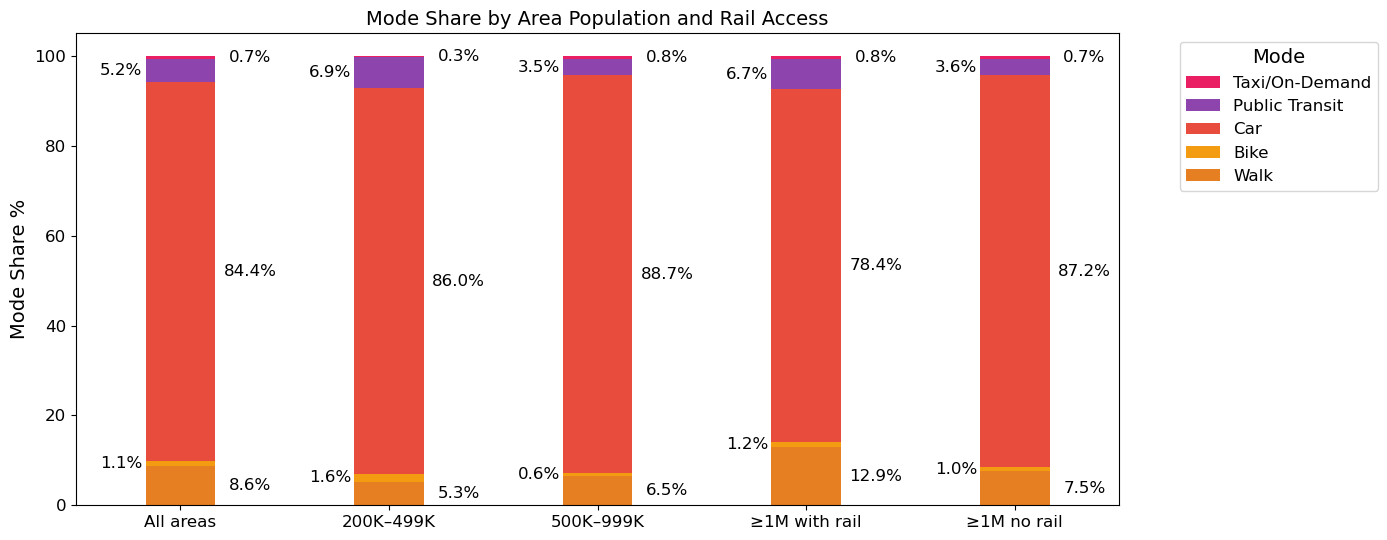}
\caption{Baseline mode shares by urban area category, reweighted from NHTS bikeable trips. Areas differ by population and rail access.}
\label{fig:mode_share_by_area}
\end{figure}

Figure \ref{fig:mode_share_by_area} shows the baseline mode share distribution in each urban typology. While car use dominates across all areas, the prevalence of non-auto modes varies substantially. Walking reaches its highest level in large cities (\emph{1M+}) with rail access (12.9\%), while public transit share is highest in small (\emph{200–499k}) cities (6.9 \%). The highest car share, instead, occurs in mid-sized (\emph{500–999k}) cities (88.7 \%)

\begin{figure}[htbp]
\centering
\includegraphics[width=0.8\textwidth]{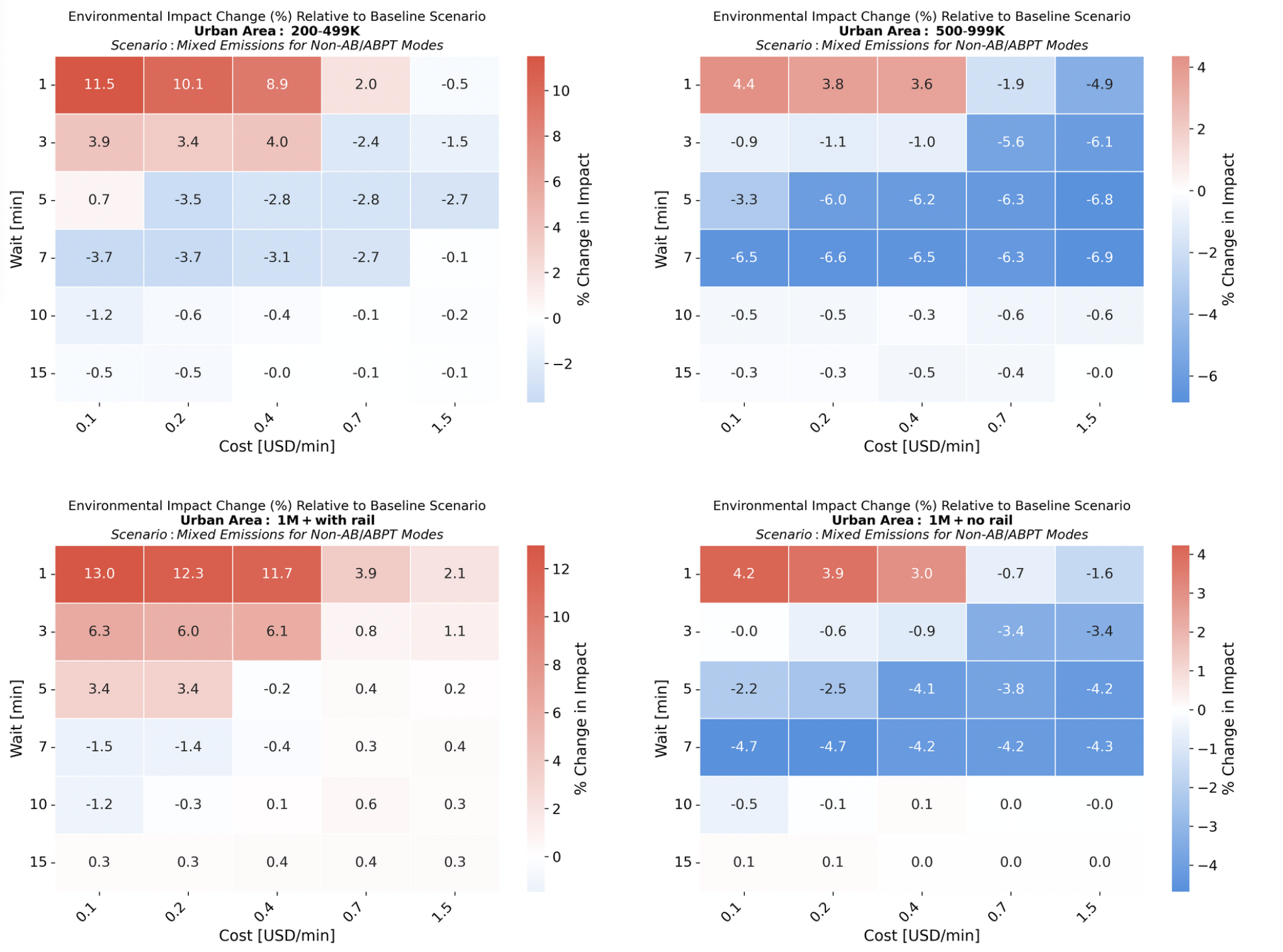}
\caption{Relative change in environmental impact (\%) compared to the baseline scenario across cost and wait time scenarios, under a Mixed Emissions background. Each subplot reflects one urban area typology.}
\label{fig:urban_env_impact}
\end{figure}

Figure \ref{fig:urban_env_impact} presents the percentage change in environmental impact relative to the baseline, considering the mixed background emissions scenario and baseline emissions for autonomous bicycles, for each urban area type and across all wait and cost combinations. Results show to be highly sensitive to baseline mobility patterns and differ notably across contexts.

Large cities with rail show the largest emissions increases—up to 13.0\% under low wait and low cost due to mode substitution from already-sustainable modes. Similarly, in smaller cities, while some scenarios lead to significant emissions reductions, the emissions increases can be large in low-wait and low-cost scenarios. Despite having relatively low walk shares, these cities exhibit the highest public transit use among the four typologies, which may help explain this result. 

In contrast, midsize cities demonstrate the greatest emissions reduction potential, reaching up to a 6.9\% emissions reduction under moderate wait and low cost. These types of urban areas have the highest car mode share, which means they also have a greater room for emissions savings via car trip replacement. Similarly, large cities without rail, where car dependence is high and public transit share is lower, also show promising reductions. This also seems to support the idea that a greater car dominance may yield higher potential for emissions improvements.

These patterns suggest that, at least in the context of mid-sized and large cities, cities with higher car dependence may offer the greatest opportunity for autonomous bicycles to reduce environmental impact, though these same areas may also face infrastructure or safety barriers to adoption. Our study does not explicitly account for perceived or actual safety, which remains a key factor in cycling behavior and should be a focus of future work.

Overall, the findings highlight that the environmental consequences of the adoption of autonomous bicycles are highly contingent on local context. A system design that yields benefits in one area may produce undesired emissions increases in another. These results underscore the need for context-specific planning and policy to ensure shared autonomous micro-mobility deployment supports—not undermines—urban sustainability goals.

\section{Discussion}\label{sec:discussion}

The impacts of emerging mobility technologies are frequently assessed only after they have been widely deployed, at which point regulatory intervention becomes more difficult. However, early-stage analysis, while inherently uncertain, is essential to guide development and inform policy before commercial and operational practices become entrenched.

This study demonstrates that critical outcomes, such as environmental impacts and adoption rates, are highly sensitive to service attributes including wait times and costs. These attributes are, to varying degrees, within the influence of public authorities. For instance, by setting fleet caps or regulating vehicle speeds, governments can indirectly shape average wait times. This relationship has been examined in prior work \cite{sanchez2022performance}. In terms of pricing, although governments may not set prices directly, they can influence cost structures through mechanisms such as per-trip fees, subsidies, or tax incentives.

The analysis also highlights demographic disparities in adoption, particularly by gender and age. These findings suggest that shared autonomous micro-mobility services may not be equitably accessed across population groups without targeted interventions. Public agencies may address these disparities by requiring operators to implement inclusive service strategies or meet predefined equity benchmarks.

Furthermore, the results underscore the context-dependent nature of impacts. This has two main policy implications. First, there is a need for location-specific studies that reflect local travel behavior, mode shares, and background fleet emissions. Second, regulatory approaches should be designed to adapt over time, recognizing that mobility needs and urban conditions will evolve. Rigid, one-size-fits-all policies are unlikely to be effective across diverse geographic and temporal contexts.

To support such adaptive and evidence-based policymaking, agencies must have access to appropriate technical expertise, either internally or through collaboration with external research institutions. Additionally, policies that mandate data sharing and performance reporting from operators are essential for validating projected impacts and ensuring ongoing accountability.

\section{Limitations}\label{sec:limitations}

As with many studies involving emerging technologies, this work is subject to potential hypothetical bias due to limited public familiarity with autonomous bicycles. To mitigate this, the survey included descriptive text, images, and examples to help respondents form a consistent understanding of the concept, as described in Section \ref{sec:survey-design}. Nonetheless, stated preference methods provide valuable early insights into potential adoption patterns and impacts. Future research should incorporate revealed preference data once such services are deployed.

Given the novelty of the technology, we sought to minimize respondent burden and maintain clarity by focusing the SP scenarios exclusively on cost and wait time. Although important factors such as safety, infrastructure availability, and social acceptability were not included directly in the experimental design, they were partially captured through latent attitudinal variables within the Hybrid Choice Model. These constructs reflect respondents’ perceptions and general openness toward shared and autonomous mobility. Future studies should aim to incorporate such attributes more explicitly, either through richer experimental designs or by integrating external datasets.

Finally, it is important to note that this study was designed specifically for a U.S. context, with all survey responses collected from U.S.-based participants. As such, the findings may not generalize to settings with different transportation systems, regulatory frameworks, or cultural attitudes toward emerging mobility technologies. Further research is needed to understand how these dynamics vary across different geographic and cultural contexts.

\section{Conclusions}\label{sec:Conclusion}

This study contributes to the field of shared autonomous micro-mobility by addressing critical gaps in understanding mode share potential, substitution patterns, environmental impacts, and demographic dimensions of adoption. Using a context-aware stated preference survey and a set of discrete choice models, including a hybrid choice model capturing latent attitudes, we simulate adoption outcomes under a range of service scenarios. Through reweighting, our results are made generalizable to broader urban populations with public transit access in the United States.

Key findings include a wide range of potential market penetration, with mode share reaching up to 23\% in the most attractive scenario, characterized by the lowest wait times and cost. However, adoption decreases sharply as either of these attributes increases. These variations also substantially affect mode substitution patterns, with important consequences for environmental outcomes.

Our analysis reveals that the scenarios with the shortest wait times tend to increase environmental impacts, primarily due to a higher shift from already-sustainable modes such as walking and biking. In contrast, more moderate wait times can yield emissions reductions by retaining a meaningful share of users shifting from private car use. Environmental outcomes also depend significantly on contextual factors, including city type, the emissions profile of the background vehicle fleet, and assumptions about the lifespan and infrastructure demands of the shared autonomous micro-mobility system. Estimated impacts range from a 6.5\% reduction to a 32\% increase in emissions, highlighting the need for context-sensitive assessments and adaptive policymaking. For instance, one could imagine that if market incentives drive operators toward low-cost, low-wait-time designs, environmental impacts could be very negative. This underscores the importance of a strong public sector role in guiding deployment toward societal goals.

From an equity perspective, our analysis did not find increased preference biases associated with higher income or education, diverging from patterns observed in traditional bikeshare systems. However, we do identify lower adoption likelihood among women and older adults—groups whose needs must be explicitly addressed in vehicle, service, and infrastructure design. Moreover, even in the absence of preference-based disparities, inequities may arise if deployments are concentrated in higher-income or otherwise segregated neighborhoods, thereby limiting access for underserved populations.

Overall, this study offers a comprehensive behavioral and environmental evaluation of shared autonomous micromobility systems. Our findings can inform more equitable, effective, and sustainable policy and implementation strategies for cities exploring new forms of lightweight autonomous transportation.

\section*{Declaration of generative AI and AI-assisted technologies in the writing process}

During the preparation of this work, the authors used ChatGPT in order to improve readability and language (i.e., for polishing text written by the authors). After using this service, the authors reviewed and edited the content as needed and take full responsibility for the content of the publication.
In particular, authors certify they reviewed the changes to make sure there was no alteration in the meaning of the text.

\section*{Appendix}
\appendix

\subsection{Indicators for attitudinal section of the survey}\label{app:Indicators}

\begin{table}[ht]
\centering
\footnotesize
\setlength{\tabcolsep}{6pt}
\renewcommand{\arraystretch}{1.2}
\caption{Indicator questions used in the attitudinal section of the survey.}
\label{tab:indicators}
\begin{tabular}{ll}
\toprule
\textbf{Indicator} & \textbf{Description} \\
\midrule
I1 & Biking is an enjoyable way to travel and a good form of exercise. \\
I2 & I would consider biking only when the weather is favorable. \\
I3 & Biking feels safe in my city, and there is adequate bike lane coverage. \\
I4 & Public transportation is efficient and reliable. \\
I5 & I know how to navigate to my destinations using public transportation. \\
I6 & It is important to use more sustainable modes of transportation. \\
I7 & My daily actions can help minimize global warming. \\
I8 & Technological innovations enhance our quality of life. \\
I9 & I enjoy being an early adopter of new technologies. \\
I10 & Autonomous bicycles would be both convenient and fun. \\
I11 & I would feel uneasy about encountering autonomous bicycles in the streets. \\
\bottomrule
\end{tabular}
\end{table}

\subsection{Binary Discrete Choice Model for Trip Bikeability}\label{app:bikeability-model}

The model specification includes individual sociodemographic attributes (e.g., gender, age, student status), trip characteristics (e.g., mode, travel time, purpose), and contextual variables (e.g., weather conditions). Car was set as the reference travel mode and "work" as the reference purpose.

The utility of perceiving a trip as bikeable is defined as follows:
\begin{align*}
V &= ASC 
+ B_{\text{walk}} \cdot (\text{mode} = \text{walk})
+ B_{\text{PT}} \cdot (\text{mode} = \text{transit})
+ B_{\text{taxi}} \cdot (\text{mode} = \text{taxi})\\
&\quad + B_{\text{time}} \cdot \mathit{time} 
+ B_{\text{fulltime}} \cdot \mathit{full\_time} 
+ B_{\text{woman}} \cdot \mathit{Sex} 
+ B_{\text{older}} \cdot \mathit{older} \\
&\quad + B_{\text{student}} \cdot \mathit{student} 
+ B_{\text{higher\_ed}} \cdot \mathit{higher\_ed} 
+ B_{\text{children}} \cdot \mathit{Children} \\
&\quad + B_{\text{time\_leisure}} \cdot \mathit{time} \cdot (\text{purpose} = \text{leisure})
+ B_{\text{harshwinter}} \cdot \mathit{harsh\_winter}
\end{align*}

Table \ref{tab:bikeability_model_results} presents the estimated coefficients, robust standard errors, and model summary statistics. The model achieves an in-sample predictive accuracy of 68.8\%, indicating good fit for classifying trips as bikeable

\begin{table}[H]
\centering
\scriptsize
\setlength{\tabcolsep}{6pt}
\caption{Model estimation results for trip bikeability ($*p < 0.1$, $**p < 0.05$, $***p < 0.01$)}
\label{tab:bikeability_model_results}
\begin{tabular}{lcc}
\toprule
\textbf{Name} & \textbf{Coefficient} & \textbf{Rob. Std Err} \\
\midrule
ASC & 1.370*** & 0.113 \\
B\_PT & 0.691*** & 0.083 \\
B\_children & 0.100* & 0.059 \\
B\_fulltime & 0.147** & 0.071 \\
B\_harshwinter & 0.154** & 0.073 \\
B\_higher\_ed & -0.123** & 0.062 \\
B\_older & -0.262*** & 0.083 \\
B\_student & 0.239*** & 0.083 \\
B\_taxi & 0.389*** & 0.070 \\
B\_time & -0.540*** & 0.019 \\
B\_time\_leisure & 0.164*** & 0.015 \\
B\_walk & 0.649*** & 0.085 \\
B\_woman & -0.271*** & 0.055 \\
\midrule
\textbf{Model Summary} & & \\
Estimated Parameters & 13 & -- \\
Sample Size & 6,117 & -- \\
Initial Log-Likelihood & -4239.98 & -- \\
Final Log-Likelihood & -3677.47 & -- \\
Likelihood Ratio Test & 1125.03 & -- \\
Rho-Square & 0.133 & -- \\
Rho-Square-Bar & 0.130 & -- \\
AIC & 7380.93 & -- \\
BIC & 7468.27 & -- \\
\bottomrule
\end{tabular}
\end{table}

\begin{table}[H]
\centering
\scriptsize
\caption{Marginal distributions before and after IPF weighting }
\label{tab:ipf_margins}
\begin{tabular}{lcccc}
\toprule
\makecell{\textbf{Variable}} & 
\makecell{\textbf{Category}} & 
\makecell{\textbf{Survey}\\\textbf{(Unweighted)}} & 
\makecell{\textbf{Survey}\\\textbf{(Weighted)}} & 
\makecell{\textbf{NHTS}\\\textbf{(Bikeable)}} \\
\midrule
\multirow{3}{*}{Trip Purpose}
& Work/School (0) & 0.343 & 0.327 & 0.327 \\
& Leisure (1)      & 0.287 & 0.317 & 0.317 \\
& Errands (2)      & 0.371 & 0.357 & 0.357 \\
\midrule
\multirow{5}{*}{Mode}
& Car (0)          & 0.194 & 0.134 & 0.134 \\
& Walk (1)         & 0.157 & 0.012 & 0.012 \\
& Bike (2)         & 0.227 & 0.785 & 0.785 \\
& Public Transit (3) & 0.191 & 0.060 & 0.060 \\
& Taxi/On-Demand (4) & 0.230 & 0.007 & 0.007 \\
\midrule
\multirow{2}{*}{Sex}
& Male (0)         & 0.507 & 0.596 & 0.596 \\
& Female (1)       & 0.493 & 0.404 & 0.404 \\
\midrule
\multirow{2}{*}{Age}
& Young (1)        & 0.282 & 0.391 & 0.391 \\
& Older (1)        & 0.149 & 0.211 & 0.211 \\
\midrule
\multirow{2}{*}{Education}
& Higher Ed (1)    & 0.694 & 0.310 & 0.310 \\
& No Higher Ed (0) & 0.306 & 0.690 & 0.690 \\
\midrule
\multirow{2}{*}{Income}
& Low Income (1)   & 0.406 & 0.222 & 0.222 \\
& High Income (1)  & 0.174 & 0.518 & 0.518 \\
\bottomrule
\end{tabular}
\end{table}

\subsection{Variable Descriptions}

\begin{table}[H]
\centering
\scriptsize
\caption{Variable Descriptions Used in Model Specification. AB stands for autonomous bicycles and ABPT stands for autonomous bicycles combined with public transit.}
\label{app:variables}
\begin{tabular}{p{1.8cm} p{3.2cm} p{4cm} p{2cm}}
\toprule
\textbf{Category} & \textbf{Variable} & \textbf{Description} & \textbf{Type / Units} \\
\midrule

\textbf{Travel Costs and Times} 
& $Cost_i$ & Travel cost of mode $i$ (scaled to \$10 USD). & Continuous \\
& $Time_i$ & Travel time of mode $i$ (scaled to hours). & Continuous \\
& $BikingTime\_scaled$ & Biking time (scaled). Used in AB, ABPT, and bike utilities. & Continuous \\
& $WalkTime\_scaled$ & Walking time (scaled). Used in walk utility. & Continuous \\
& $CarTime\_scaled$ & Car travel time (scaled). & Continuous \\
& $PTCost\_scaled$ & Public transit cost (scaled). & Continuous \\
& $ABPT\_TotalCost\_scaled$ & Total cost of ABPT trip (scaled). & Continuous \\
& $ABPT\_BikeTime\_scaled$ & Bike portion of ABPT time (scaled). & Continuous \\
& $ABPT\_TotalTime\_scaled$ & Total time of ABPT trip (scaled). & Continuous \\

\midrule
\textbf{Wait Times} 
& $ab\_wait\_time\_scaled$ & Wait time for AB and ABPT (scaled to hours). & Continuous \\
& $pt\_shortWait$ & PT wait time < 10 minutes. & Dummy (0/1) \\
& $TaxiWaitTime\_scaled$ & Wait time for taxi (scaled to hours). & Continuous \\

\midrule
\textbf{Sociodem. Variables} 
& $high\_income$ & Income over \$100k/year. & Dummy (0/1) \\
& $low\_income$ & Income under \$35k/year. & Dummy (0/1) \\
& $full\_time$ & Full-time employment. & Dummy (0/1) \\
& $higher\_ed$ & Has university-level education. & Dummy (0/1) \\
& $children$ & Presence of children in household. & Dummy (0/1) \\
& $car\_owner$ & Car ownership. & Dummy (0/1) \\
& $white$ & Respondent identifies as white. & Dummy (0/1) \\
& $sex$ & Self-reported sex. & Categorical \\
& $woman$ & Respondent self-reported as female. & Dummy (0/1) \\
& $older$ & Age > 50. & Dummy (0/1) \\
& $young$ & Age < 30. & Dummy (0/1) \\
& $student$ & Currently enrolled as a student. & Dummy (0/1) \\
& $hot\_summer$ & Lives in top 10 hottest US states. & Dummy (0/1) \\
& $harsh\_winter$ & Lives in state with 50+ in snowfall annually. & Dummy (0/1) \\

\midrule
\textbf{Trip-Specific Variables} 
& $work\_trip$ & Trip made for work. & Dummy (0/1) \\
& $leisure\_trip$ & Trip made for leisure. & Dummy (0/1) \\
& $errands\_trip$ & Trip made for errands. & Dummy (0/1) \\

\bottomrule
\end{tabular} 
\end{table}

\subsection{Model Parameters}
 \begin{table}[H]
\centering
\scriptsize
\caption{Model Parameters and Latent Variable Components}
\label{app:parameters}
\setlength{\tabcolsep}{4pt} 
\renewcommand{\arraystretch}{1.15} 
\begin{tabular}{>{\raggedright\arraybackslash}p{2.2cm} >{\raggedright\arraybackslash}p{2.2cm} >{\raggedright\arraybackslash}p{4.6cm} >{\raggedright\arraybackslash}p{2cm}}
\toprule
\textbf{Category} & \textbf{Parameter} & \textbf{Description} & \textbf{Type} \\
\midrule

\textbf{Alternative-Specific Constants (ASCs)} 
& $ASC_i$ & Alternative-specific constant for mode $i$ (car normalized to 0). & Estimated (fixed or random) \\

\midrule
\textbf{Utility Coefficients} 
& $B_{\text{cost}}$ & Coefficient for travel cost. & Estimated (fixed or random) \\
& $B_{\text{time}}$ & Coefficient for travel time. & Estimated (fixed or random) \\
& $B_{\text{activetime}}$ & Coefficient for active travel time (walk/bike). & Estimated (fixed or random) \\
& $B_{\text{wait}}$ & Coefficient for wait time. & Estimated (fixed) \\
& $B_{\text{ptshortwait}}$ & Coefficient for short PT wait. & Estimated (fixed) \\
& $B_x$ & Coefficients for trip-specific and sociodemographic dummies. & Estimated (fixed) \\

\midrule
\textbf{Latent Variable Specification (Model 3)} 
& $LV_{\text{AB}}$ & Latent variable for negative attitude toward ABs. & Latent (continuous) \\
& $B_{\text{LV}}$ & Effect of $LV_{\text{AB}}$ in AB and ABPT utility (tanh-transformed). & Estimated (fixed) \\
& $\text{coef}_x$ & Structural equation coefficients for sociodemographics. & Estimated (fixed) \\
& $\sigma_s$ & Standard deviation of latent variable's error term. & Estimated (random) \\
& $\omega$ & Random draw for latent variable simulation. & Simulated (standard normal) \\

\midrule
\textbf{Measurement Model (Indicators)} 
& $I10$, $I11$ & Indicators measuring attitudes toward ABs. I10 and related params fixed for scale and identification. & Observed (ordinal) \\
& $B_{I11}$ & Loading of $LV_{\text{AB}}$ on I11. & Estimated \\
& $INTER_{I11}$ & Intercept for I11 (ordered probit). & Estimated \\
& $SIGMA_{I11}$ & Scale parameter for I11. & Estimated \\

\bottomrule
\end{tabular}
\end{table}

\subsection{Model evaluation: Cross-validation accuracy and error in mode share prediction} \label{app:cv-tables}

\begin{table}[H]
\centering
\footnotesize
\caption{Cross-validation accuracy by mode for Models 1–3 (Mean ± Std) using five folds. AB stands for autonomous bicycles and ABPT stands for autonomous bicycles in combination with public transit.}
\begin{tabular}{lccc}
\toprule
\textbf{Mode / Metric} & \textbf{Model 1} & \textbf{Model 2} & \textbf{Model 3} \\
\midrule
Mode 0 (Walk)        & 0.920 $\pm$ 0.067 & 0.750 $\pm$ 0.071 & 0.925 $\pm$ 0.053 \\
Mode 1 (Bike)        & 0.974 $\pm$ 0.041 & 0.977 $\pm$ 0.036 & 0.978 $\pm$ 0.037 \\
Mode 2 (Car)         & 0.866 $\pm$ 0.054 & 0.793 $\pm$ 0.080 & 0.854 $\pm$ 0.067 \\
Mode 3 (Transit)     & 0.697 $\pm$ 0.086 & 0.718 $\pm$ 0.059 & 0.698 $\pm$ 0.100 \\
Mode 4 (Taxi)        & 0.672 $\pm$ 0.088 & 0.639 $\pm$ 0.088 & 0.737 $\pm$ 0.065 \\
Mode 5 (AB)       & 0.466 $\pm$ 0.157 & 0.611 $\pm$ 0.112 & 0.432 $\pm$ 0.170 \\
Mode 6 (ABPT)     & 0.265 $\pm$ 0.082 & 0.236 $\pm$ 0.066 & 0.315 $\pm$ 0.109 \\
Mean Accuracy        & 0.653 $\pm$ 0.024 & 0.644 $\pm$ 0.021 & 0.664 $\pm$ 0.017 \\
Std Dev (per mode)   & 0.257 $\pm$ 0.038 & 0.222 $\pm$ 0.023 & 0.254 $\pm$ 0.030 \\
\bottomrule
\end{tabular}
\end{table}

\begin{table}[ht]
\centering
\footnotesize
\caption{Mean Absolute Difference (\%) in mode share prediction for Models 1–3 (Mean ± Std Dev) using cross-validation. AB stands for autonomous bicycles and ABPT stands for autonomous bicycles in combination with public transit.}
\begin{tabular}{lccc}
\toprule
\textbf{Mode} & \textbf{Model 1} & \textbf{Model 2} & \textbf{Model 3} \\
\midrule
Walk (0)      & 2.39 $\pm$ 1.76 & 2.04 $\pm$ 2.13 & 2.49 $\pm$ 1.73 \\
Bike (1)      & 1.68 $\pm$ 0.82 & 1.72 $\pm$ 0.72 & 1.70 $\pm$ 0.72 \\
Car (2)       & 1.26 $\pm$ 1.56 & 1.72 $\pm$ 2.45 & 1.18 $\pm$ 1.74 \\
Transit (3)   & 1.27 $\pm$ 0.97 & 1.00 $\pm$ 1.11 & 1.11 $\pm$ 1.03 \\
Taxi (4)      & 2.15 $\pm$ 2.22 & 2.76 $\pm$ 1.98 & 1.57 $\pm$ 1.71 \\
AB (5)     & 3.14 $\pm$ 4.11 & 7.85 $\pm$ 4.36 & 3.85 $\pm$ 4.44 \\
ABPT (6)   & 3.15 $\pm$ 1.02 & 3.91 $\pm$ 2.63 & 2.29 $\pm$ 0.81 \\
\bottomrule
\end{tabular}
\end{table}

\subsection{Origin modes for new autonomous bicycle users}

\begin{figure}[H]
\centering
\includegraphics[width=\textwidth]{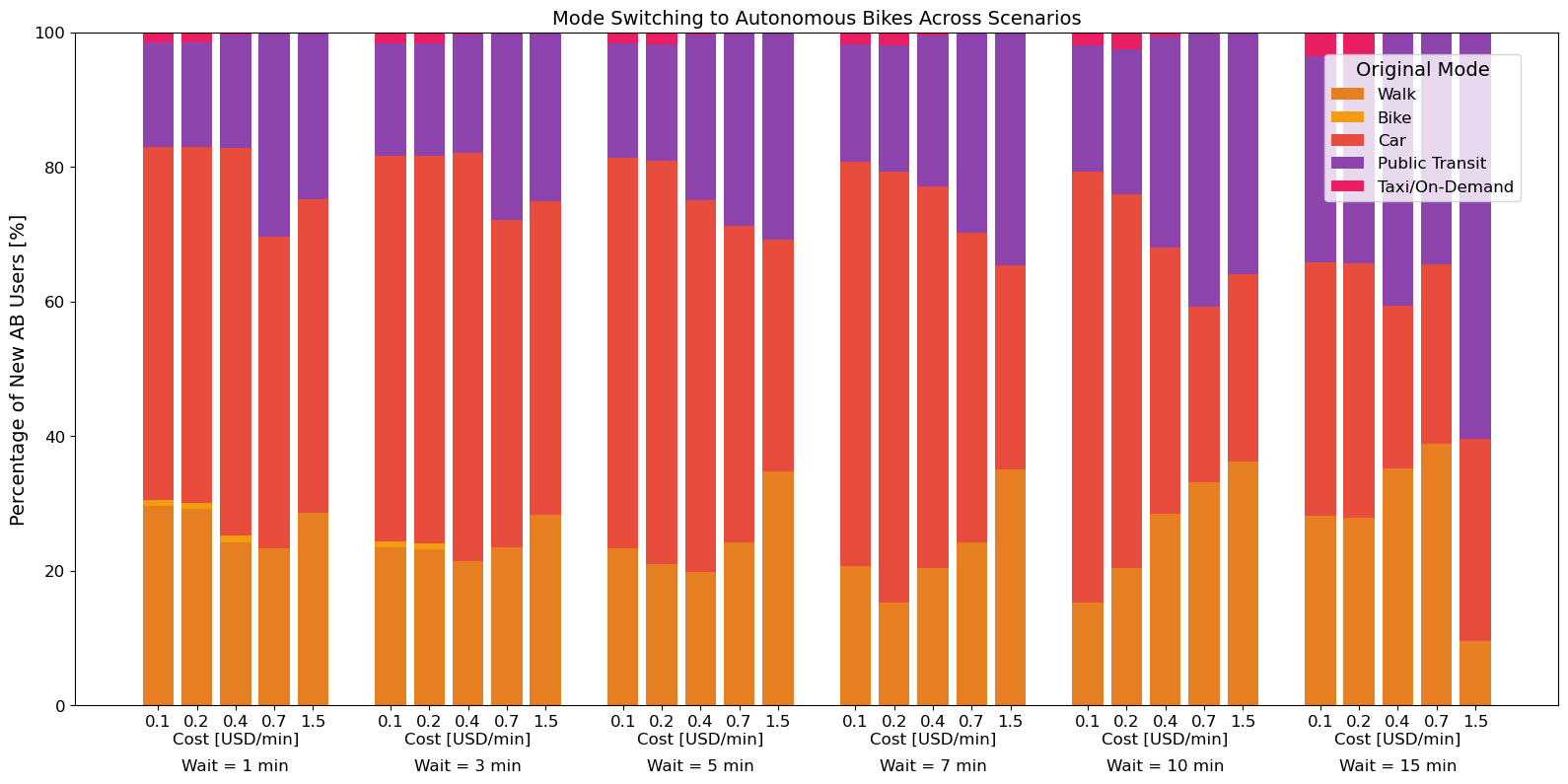}
\caption{Distribution of original modes among new autonomous bike users across all modeled wait time and cost scenarios.}
\label{fig:ab_shift}
\end{figure}

\begin{figure}[H]
\centering
\includegraphics[width=\textwidth]{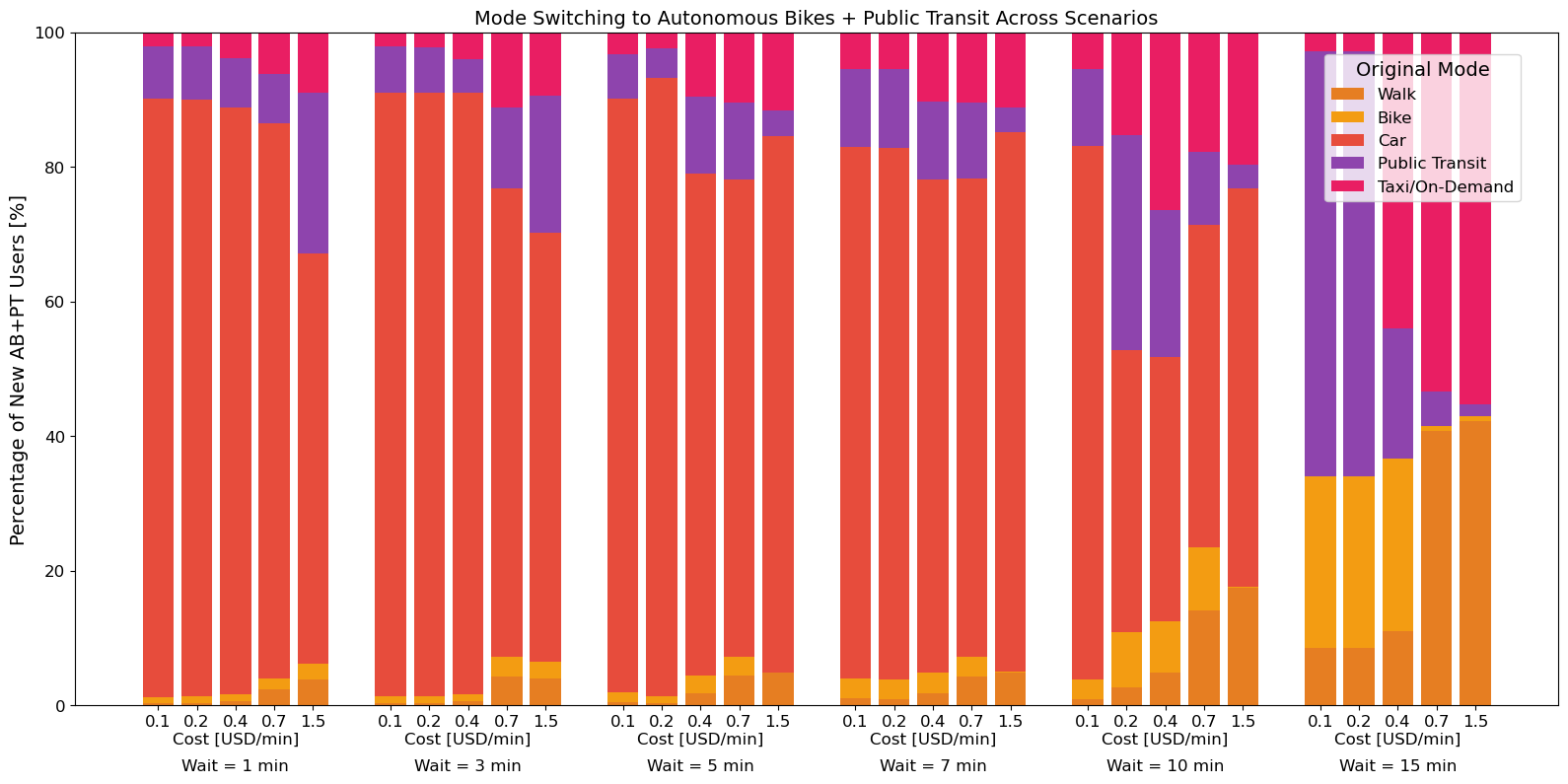}
\caption{Distribution of original modes among new users of autonomous bikes combined with public transit across all modeled wait time and cost scenarios.}
\label{fig:abpt_shift}
\end{figure}

\bibliographystyle{elsarticle-harv} 
\bibliography{cas-refs}

\end{document}